\documentstyle[11pt,aaspp4,psfig,lscape]{article}

\def\rmit#1{#1}

\def\vs{\rmit{vs.}}

\def\specchar#1{{\sc #1}}

\def\FeI{\mbox{Fe\,\specchar{i}}}
\def\FeII{\mbox{Fe\,\specchar{ii}}}

\def\[OI]{\mbox{[O\,\specchar{i}]}}

\def\degree{\hbox{$^\circ$}}

\def\kms{\hbox{km$\;$s$^{-1}$}}

\begin{document}

\title{Oxygen in the Very Early Galaxy\footnote{Based on data collected at the
Keck\,I, VLT and William Herschel telescopes}}
\author{Garik Israelian, Rafael Rebolo\altaffilmark{2}, Ram\'on J. Garc\'\i a L\'opez\altaffilmark{3}} 
\affil{Instituto de Astrof\'\i sica de Canarias, E-38200
La Laguna, Tenerife, Spain}
\author{Piercarlo Bonifacio, Paolo Molaro}
\affil{Osservatorio Astronomico di Trieste, Via G. B. Tiepolo 11, I-34131 Trieste, Italy} 
\author{Gibor Basri}
\affil{Astronomy Department, University of California,
Berkeley, California 94720, USA}
\begin{center}
and
\end{center}
\author{Nataliya~Shchukina}
\affil{Main Astronomical Observatory, National Academy of Sciences,
03680~Kyiv-127, Ukraine}
\authoraddr{Instituto de Astrof\'\i sica de Canarias, E-38200 La Laguna, 
Tenerife, SPAIN}

\altaffiltext{2}{Consejo Superior de Investigaciones Cient\'\i fcas, Spain}
\altaffiltext{3}{Departamento de Astrof\'\i sica, Universidad de La Laguna, 
E-38071 La Laguna, Tenerife, Spain}

\authoremail{gil@ll.iac.es, rrl@ll.iac.es, basri@soleil.Berkeley.edu, 
rgl@ll.iac.es, shcukina@iac.es}

\newpage

\begin{abstract}

Oxygen abundances in a sample of ultra-metal-poor subdwarfs have been 
derived from measurements of the oxygen triplet at 7771--5 \AA\ and OH lines in 
the near UV performed in high-resolution and high signal-to-noise spectra 
obtained with WHT/UES, KeckI/HIRES, and VLT/UVES.   Our Fe abundances were 
derived in LTE and then corrected for 
NLTE effects following Th\'evenin \& Idiart (1999). The new oxygen 
abundances confirm previous findings for a progressive linear rise in the 
oxygen-to-iron ratio with a slope $-$0.33$\pm$0.02 from solar
metallicity to [Fe/H]$\sim -$3. A slightly higher slope would be 
obtained if the Fe NLTE corrections were not considered. 
Below [Fe/H]= $-$2.5 our stars show [O/Fe] ratios as high as $\sim$1.17 
(G64-12), which can be interpreted as evidence for oxygen overproduction 
in the very early epoch of the  formation of the halo, possibly 
associated with supernova events with very massive progenitor stars. 
We show that the arguments against this linear trend given by Fulbright
\& Kraft (1999), based on the LTE Fe analysis of two metal-poor stars cannot 
be sustained when an NLTE analysis is performed. We discuss how the
Fulbright \& Kraft (1999) LTE ionization balance of Fe lines underestimates 
the gravity of the very metal-poor star BD +23\arcdeg3130 
([Fe/H]=-2.43) and how this leads to an underestimation 
of the oxygen abundance derived from the forbidden line. Gravities from 
{\it Hipparcos} appear to be in good agreement with those determined in NLTE, 
giving higher values than previously assumed, which reduces the discrepancies
between the oxygen abundances determined from OH, triplet, and forbidden lines.
Using 1-D models our analysis of three oxygen indicators available for 
BD +23\arcdeg3130 gives consistent abundances within 0.16 dex and average
[O/Fe] ratio of 0.91. The high oxygen abundances at very low metallicities 
do not pose a problem to theoretical modeling since there is a range of
parameters in the calculations  of nucleosynthesis yields from massive stars at
low metallicities that  can accommodate our results.

\end{abstract}


\newpage

\section{Introduction}

The halo dwarfs contain fossilized records of the early Galaxy's composition. 
The chemical composition of their atmospheres is not altered by any internal mixing 
and provides a unique opportunity to constrain models of the chemical evolution of 
the Galaxy. Oxygen is a key element in this scheme. It is produced in the interiors
of massive stars by hydrostatic burning and returned to the interstellar medium (ISM)
when the stars explode as Type II supernovae. 

There have been numerous investigations of the oxygen abundances in halo stars.
Despite considerable observational effort the trend of the [O/Fe], ([O/Fe]=log
(O/Fe)$_\star$--log (O/Fe)$_\odot$), ratio in the halo is still unclear. Analysis 
of the forbidden line [O\,{\sc i}] 6300 \AA\ in giants shows a $plateau$ with
[O/Fe] = 0.4--0.5 in the metallicity domain $-3<$ [Fe/H] $<$ 0 (\cite{bar88}; 
\cite{kraf92}). On the other hand, results based on the near-IR 
triplet at 7771--5 \AA\ in unevolved subdwarfs (Abia \& Rebolo 1989; \cite{tom92};
\cite{kin95}; Cavallo, Pilachowski, \& Rebolo 1997) point towards increasing 
[O/Fe] values with decreasing [Fe/H], reaching a ratio $\sim 1$ for stars with 
[Fe/H] $\sim -3$ (but see Carretta, Gratton, \& Sneden 2000) and suggesting a 
higher production of oxygen during the early life of the Galaxy. More recent NLTE
analyses of the triplet in metal-poor stars have been performed by 
Mishenina et al. (2000), Carretta et al. (2000), and Takeda et al. (2000)
showing a range for the NLTE abundance corrections.

Studies based on OH lines in the near-UV (Bessell, Sutherland, \& Ruan 1991; 
Nissen et al. 1994) concluded that [O/Fe] is constant at [Fe/H] $< -1$, 
in agreement with results based on the forbidden lines in giants. Contrary 
to this, oxygen abundances derived from near-UV OH lines for 24 
metal-poor stars by Israelian, Garc\'\i a L\'opez, \& Rebolo (1998) 
show that the [O/Fe] ratio increases from 0.6 to 1 between
[Fe/H] = $-$1.5 and $-3$ with a slope of $-0.31 \pm$ 0.11. These
new abundances derived from low-excitation OH lines agreed well with
those derived from high-excitation lines of the O\,{\sc i} triplet.
The comparison with oxygen abundances derived using O\,{\sc i} data from
Tomkin et al. (1992) showed a mean difference of $0.00\pm 0.11$ dex for the
stars in common. Boesgaard et al. (1999a), using high-quality
 Keck spectra of many metal-poor stars in the near UV, found a very good
agreement with the results obtained by Israelian et al. (1998), and  basically
the same dependence of [O/Fe] versus metallicity. The mean difference in oxygen
abundance for ten stars in common is $0.00\pm 0.06$ dex when the differences in
stellar parameters are taken into account.  

The linear trend of the [O/Fe] ratio found in the analysis of OH and triplet
lines has been recently questioned by Fulbright \& Kraft (1999) (hereafter FK)
who obtained low values of [O/Fe] for two metal-poor subgiants using 
the [O\,{\sc i}] lines. In the present paper we discuss their claims in the 
light of a NLTE analysis of Fe lines and report on new detections of the OH 
and O\,{\sc i} triplet lines in several unevolved ultra-metal-poor stars 
with [Fe/H] $\leq -$2.5.

\section{The Observations}

The spectroscopic observations of several metal-poor subdwarfs were 
carried out in different runs using the 4.2 m WHT/UES (La Palma), the
10 m KeckI/HIRES (Hawaii), and the 8.2m VLT Kueyen/UVES (Paranal). See the
observing log in Table 1. 
All observations with WHT/UES were obtained using the E31 grating. 
In the first run (1996) we observed  BD +03\arcdeg740, BD +23\arcdeg3130, 
and G4-37 using a 1\farcs{1} arcsec slit  and a TEK CCD 
(1024 $\times$ 1024 pixel$^{2}$ giving a dispersion of 0.06 \AA\ pix$^{-1}$). 
The same configuration  but a CCD SITE1  (2148 $\times$ 2148 pixel$^{2}$ detector 
giving a dispersion of  0.076 \AA\ pix$^{-1}$)
was used to obtain a few images of G64-37 in 1997. This star, together with 
G268-32, was  observed in 1999 with the same configuration and a slit width
of  3\farcs{0}. All the
WHT/UES images were reduced using standard IRAF routines. Normalized 
flats created for each observing night were used to correct the pixel-to-pixel
variations and a ThAr lamp was used to find a dispersion solution. 
Observations with the 2.5 m  Nordic Optical 
Telecope (NOT) and the SOFIN high-resolution spectrograph were used only 
to study the Fe abundance of some of the targets. A resolving power of
$\sim$ 30\,000 has been achieved with the short camera (350 mm focal length, third 
camera) with a slit width of $\sim$1\farcs{0}. The CCD used in this run was
a $1152 \times 770$ pixel$^{2}$ EEV.

The Keck observations were carried out with the HIRES spectrograph and the TEK
2048  $\times$ 2048 pixel$^{2}$ CCD, which has been its only detector to date.  
We did not bin the chip, so the resolution is about 60\,000 with the 
1\farcs{1} slit.  The slit length was 14\farcs{0}, allowing  good sky 
subtraction.  The seeing was slightly worse than 1\farcs{0}.  Standard
calibration exposures and procedures were used to reduce the
data.  A red wavelength setting was used to observe the oxygen triplet lines; 
this setting has incomplete coverage of the echelle format.

The UVES data for G271-162, G275-4, and LP 815-43 were obtained during
the commissioning of the instrument, while the data for G64-12  were   
obtained during the science verification.
UVES has two arms:  blue and  red. In the blue arm the detector is an 
EEV of $4096\times 2048$ square pixels of  size 15 $\rm\mu m $.
In the red arm the detector is a mosaic of two CCDs: an EEV identical
to that of the blue arm and an MIT with the same number of pixels and
pixel size as the EEV. The data for  G271-162  consist of four
spectra in the red arm centered at 7100 \AA ~ with a slit width  of 
0\farcs{3} which results in a resolution of $\approx$ 110 000. 
These spectra are the same ones used by Nissen et al. (2000)
altough they have been independently reduced by us.
The O\,{\sc i} triplet falls on the MIT CCD.
The spectra of G271-162 were trailed along the slit.
For G275-4 the data consist of three spectra taken with dichroic \#{1}, the
blue arm spectra are all centered at 3460 \AA, while the red arm spectra
are centered at 5400 \AA, 5800 \AA, and 8600 \AA. The only red 
spectrum used in this paper is that centered at 8600 \AA, in which 
the triplet (not detected) falls on the EEV CCD. The slit width 
was 1\farcs{0} both for the red and the blue arm. For LP 815-43 we also 
have three spectra, but taken in the blue arm only (no dichroic). 
Two spectra were taken with a slit width of 1\farcs{1}
and one with a slit width of 0\farcs{8}. In order to combine the spectra, 
the higher-resolution spectrum has been degraded to  lower resolution
by convolving with a Gaussian of appropriate FWHM. For G64-12 the data 
consist of four blue and four red spectra observed with dichroic \#{1}.
All blue arm spectra are centered at 3460 \AA. 
Two of the red arm spectra are centered at 5800 \AA\ and two
at 8600 \AA. The triplet falls on the EEV CCD of these latter 
frames. The slit width was always 0\farcs{8}  for both arms. 
The UVES spectra have been background-subtracted, filtered for cosmic rays,
flat--fielded, extracted, and wavelength-calibrated using the echelle 
context of {\tt MIDAS}.

Our detections of the  oxygen triplet are presented
in Figure 1. We have detected for the first time the oxygen triplet in 
G268-32, G4-37, and G271-162. We also report on a possible detection of the 
bluest line of the triplet with an equivalent width of 6 m\AA\ in G64-12. 
Unfortunately the line is located at the edge of the order (which in 
addition suffers strong fringing) and therefore we cannot give much weight 
to this detection. The presence of the triplet is confirmed in  
BD $+$03\arcdeg740 (the strongest line was already measured by Tomkin et al. 
1992) and in BD $+$23\arcdeg 3130. In the latter case we have detected all 
three lines of the triplet and measured equivalent widths of $7 \pm 2$, 
$5 \pm 2$, and $3 \pm 2$ m\AA, in good agreement with the observations 
of Mishenina et al. (2000).

We have achieved the detection of OH lines in the near-UV spectra of three
ultra-metal-poor subdwarfs observed during  VLT/UVES commissioning and science 
verification. Most of the unblended OH lines carefully selected and studied by 
Israelian et al. (1998) have been unambiguously detected in the spectra 
of G64-12, G275-4 (CD$-$24\arcdeg17504), and LP815-43. Some of these lines 
are shown in  Figure 2. The S/N in the continuum is given in Table 1. The 
quality of these spectra allows us to perform spectral synthesis and to 
determine the abundance of oxygen.

\section{Spectral synthesis, stellar parameters and abundances}

The LTE spectrum synthesis program WITA (Pavlenko 1991) and ATLAS9 model 
atmospheres (Kurucz 1992) were employed in the present study.
However, in this paper we have also performed 
computations with the MOOG code (Sneden 1973) in order to be sure that there are
no systematic differences connected with the  use of different spectrum synthesis
packages. We found that in all cases considered the agreement between WITA and 
MOOG was very good. The solar oxygen abundance used for the differential 
analysis of OH lines 
was log $\epsilon$(O)=8.93 (on the customary scale in which log $\epsilon$(H)=12) 
according to Anders \& Grevesse (1989). For the triplet and [O\,{\sc i}] 
we used the solar abundances log $\epsilon$(O)=8.98 and log $\epsilon$(O)=8.90, 
respectively, derived by Israelian et al. (1998). The oscillator strengths of 
the O\,{\sc i} triplet and iron lines were taken from Tomkin et al. (1992) and 
O'Brian et al. (1991), respectively.  Effective temperatures ($T_{\rm eff}$) 
for our stars were estimated using the Alonso, Arribas,  \& Mart\'\i nez-Roger (1996) 
and Carney (1983) calibrations versus $V$$-$$K$ colors, 
and cover a wide range  of spectral types and metal content. The $K$-band
photometry was kindly provided by A. Alonso prior the publication of the data
(A. Alonso, private communication). Our final
$T_{\rm eff}$ is an average value obtained from these two calibrations. 
Since there are no accurate {\it Hipparcos} parallaxes available for our targets
(except  for BD $+$23\arcdeg 3130) we estimated surface gravities by
comparing Str\"omgren $b-y$ and c$_1$ indices with synthetic ones generated
using the corresponding filter transmissions and a grid of ATLAS9 
blanketed model atmosphere fluxes. For BD $+$23\arcdeg 3130 we obtained
$\log g = 3.05 \pm 0.25$ from the {\it Hipparcos} parallax (Perryman et al. 1997) 
following the recipe given in Allende Prieto et al. (1999) and assuming 
an IRFM based $T_{\rm eff}$=5130$\pm$150 K from Israelian et a. (1998). 
The final stellar parameters of our targets (except BD $+$23\arcdeg 3130, 
which will be treated separately in section 5), along with a representative
selection of  parameters reported by different  authors in the literature
(by no means exhaustive), are listed in Table 2. It can be seen that our 
parameters are not very different from those reported in different sources.

For the present analysis we used the same model atmospheres than Israelian et al. (1998) 
provided by  Kurucz (1992, private communication) and built with the approximate 
overshooting option switched on (note that a typographic error mistakenly stated 
that the models used in Israelian et al. had no overshooting). The 
differences between models with and without convective overshooting have been
discussed by Castelli, Gratton \& Kurucz (1997) and Molaro, Bonifacio
\& Primas (1995). We have explicitly computed the effects of overshooting
 versus non-overshooting in  the formation
of OH, [O\,{\sc i}], oxygen triplet and the Fe lines listed in Tables 3 and 4 for
a range of model parameters, and find that the differences induced in the 
[O/Fe] ratios by the two types of models  are less than 0.05 dex. 

The LTE metallicities of our targets have been estimated using a sample of 
unblended Fe\,{\sc i} lines listed in Tables 3 and 4 with accurately known
oscillator strengths. The equivalent widths of the iron lines were measured
in the spectra listed in Table 1. For only one star, G271-162, we took the 
measurements from the literature (Zhao \& Magain 1990). NLTE estimates
for  the Fe abundances of our stars have been made using the work of
Th\'evenin \& Idiart (1999). In their Fig. 9 we can see  NLTE
corrections for Fe abundances in metal-poor stars as a function of 
[Fe/H]$_{LTE}$. In a first approximation we can see that these corrections 
mostly depend on metallicity and there is no clear dependence on the effective 
temperatures and gravities. So, we have fitted a third order polynomial to  the
data provided in their  Table 1, suitably reproducing the trend seen in their 
figure. The resulting polynomial fit is

\begin{eqnarray*}
[Fe/H]_{NLTE} - [Fe/H]_{LTE}=-0.001-0.204\times[Fe/H]_{LTE}-\\  
0.052\times([Fe/H]_{LTE})^{2}-0.006\times([Fe/H]_{LTE})^{3}
\end{eqnarray*}

We have used this function to estimate the 
[Fe/H]$_{NLTE}$ for our stars and listed the resulting values
in Table 7.  

All our targets except two (BD $+$03\arcdeg 740 and BD $+$23\arcdeg 3130) are
relatively distant, and their colors could be affected by  interstellar reddening.
 Unfortunately, the reddening values of our targets are very 
uncertain and we have decided not to consider them.  
The reddening correction  will increase 
the $T_{\rm eff}$ values of our stars which will increase the oxygen 
abundances derived from the OH and [O\,{\sc i}] lines and diminish 
the abundance derived from the triplet.

In Table 3 we present oxygen LTE abundances from the parameters given in Table 2. 
It is well known that the oxygen abundance in metal-poor subdwarfs derived from 
the O\,{\sc i} is slightly sensitive to NLTE effects (Kiselman 1993, Takeda 1995). 
Tomkin et al. (1992) estimate NLTE corrections of less than 0.1 dex for 
hot subdwarfs with  $T_{\rm eff}$ around 6200 K. Our stars have these 
temperatures, and corrections  will be of this order or smaller due to 
the weakness of our triplet lines. However, larger NLTE corrections 
could  be necessary if the hydrogen collision rates 
employed by Tomkin et al. (1992) were overestimated, as suggested 
by Fleck et al. (1991) and Belyaev et al. (1999).

Oxygen abundances in G275-4, G64-12, and LP815-43 were derived by fitting the 
OH lines (see Figure 3) in the near-UV with oscillator strengths adjusted
to reproduce the solar spectrum ($\log gf_{\rm ad}$, used by
Israelian et al. 1998). The adjustments were most probably due to
imprecise gf-values of other weak lines blended with OH. It is well
known that the line lists available in the literature are not complete 
and/or perfect and Israelian et al. (1998) tried to empirically minimize 
the effect they could have by adjusting the $gf$'s of the OH. It is possible
that some of these blended lines will disappear in metal-poor stars 
(especially if they belong to one of the Fe-group elements) causing a systematic
difference in oxygen abundance. However, this will not affect 
our [O/H] ratios by more than 0.1 dex.

The results for each star and 
each line are listed in Table 4. In this table we also provide theoretical 
$\log gf_{\rm th}$ values derived from new transition probabilities recently
computed by Gillis et al. (2000). When 
computing new theoretical $gf$ values from the transition probabilities
provided by Gillis et al. (2000), we found two mistakes in  Table 2 of 
Israelian et al.(1998). The values of $\log gf_{\rm th}$ for the lines 3167.169 
and 3186.084 \AA\ were not correct. The correct values of 
$\log gf_{\rm th}$ must be $-$1.540 and $-$1.830 for the 3167.169 and 
3186.084 \AA\ lines, respectively. It can be seen from  Table 4 and Table 2 
of Israelian et al. (1998) that these new $gf$ values of Gillis et al. (2000)
are almost identical to those of Goldman \& Gillis (1981) and that the maximum
difference between the $\log gf$ values adjusted to reproduce
the solar OH lines and the new theoretical ones is $-$0.24 dex, while the mean 
difference correction is $-$0.11 dex.   
We note that the most accurate measurement is achieved 
in G64-12 where the error in oxygen abundance due to the continuum placement 
for individual OH lines is as small as 0.2 dex (see Figure 3). However, 
in the case of G275-4 these errors may reach 0.35 dex. We regard the 
clear detection of OH lines in G64-12 as an important confirmation of the linear 
trend of [O/Fe] versus [Fe/H]. Indeed, even if we suppose that the
near-UV continuum is not well determined and that the $gf$ values of 
the OH lines are underestimated, the [O/Fe] ratio in G64-12 will decrease 
because of this effect by 0.1 at most. In addition, assuming for this 
star the smallest effective temperature reported in the literature 
($T_{\rm eff} = 6197$ K; Table 2) would bring the [O/H] ratio down by another 
0.2 dex. In any case we would obtain a value  of [O/Fe] $>$ 0.9.

\section{How reliable are the abundances derived from the O\,{\sc i} triplet,
OH, and [O\,{\sc i}] ?}

Recently, Mishenina et al. (2000) performed a NLTE analysis of the O\,{\sc i} 
triplet to re-derive oxygen abundances for a sample of 38 metal-poor
stars  from the literature. They confirmed earlier results (Abia \& Rebolo
1989; Tomkin et al. 1992; Kiselman 1993) indicating that the mean value of the
NLTE  correction in unevolved metal-poor stars is typically 0.1 dex and
never exceeds 0.2 dex. Mishenina et al. found the same linear trend as
Israelian et al. (1998) and Boesgaard et al. (1999a) from the OH lines and
confirmed that  oxygen abundances do not show any trend with $T_{\rm eff}$ or
$\log g$ (Boesgaard et al. 1999a). Furthermore, Asplund et al. (1999) showed
that the O\,{\sc i} triplet is not affected by 3D effects, convection, or
small-scale inhomogeneities in the stellar atmosphere. In addition, oxygen
abundances derived from this triplet are not significantly affected by
chromospheric activity either, showing that the O\,{\sc i}  
triplet provides reliable oxygen abundances in metal-poor dwarfs. 
Preliminary calculations by Asplund et al. (1999) and Asplund (2000) show that 
3D models give  stronger OH lines than those computed with 1D models 
which would decrease the oxygen abundances 
found in the present paper. However, first the precise value of 3D corrections 
will consistently require a revision of the stellar parameters according to 
the same models and, second, Fe abundances including NLTE effects should
be obtained with these models in order to properly estimate the [O/Fe] ratio.

Carretta et al. (2000) performed an independent 
analysis of 32 metal-poor stars hotter than 4600 K using the triplet,
and provide LTE and NLTE oxygen abundances that are significantly lower 
than those found in previous works. A preliminary attempt to understand 
the reasons for this discrepancy can be made by looking in detail into 
their most metal-poor star (BD $+$03\arcdeg740, [Fe/H] $= -2.66$), where 
a surprisingly low oxygen abundance
([O/Fe] $= 0.38$) is claimed. A recent study of stellar parameters based on the
NLTE analysis of iron lines (Th\'evenin \& Idiart 1999), gives a lower
effective temperature by 140 K (actually very similar to the temperature
we have derived in this paper) and a higher gravity by 0.3 dex (also very 
similar to ours)  than the
values adopted by Carretta et al., for this star.  Using the 
Th\'evenin \& Idiart parameters we obtain an LTE oxygen 
abundance 0.4 dex higher than that found
by Carretta et al. (i.e., [O/Fe]$_{\rm LTE}=1.05$), and for a star with 
these parameters the NLTE
correction to the oxygen abundance is of the order of 0.05 dex (Mishenina et
al. 2000; Tomkin et al. 1992), much lower than the 0.25 dex value used by 
Carretta et al. We therefore arrive at a value [O/Fe]$_{\rm NLTE}\sim 1.0$, 
in good agreement with the OH determination by Boesgaard et al. (1999a). 
The oxygen abundances derived by Carretta et al., are computed using the 
gravities from Gratton, Carretta, \& Castelli (1996), whose values come 
from an LTE Fe ionization balance. 
As a matter of fact, Gratton et al. (1996) remark that their gravities are
considerable lower (see their Figure 2) than those derived from the 
color--magnitude diagram. They also suggest that this discrepancy may be 
due to a NLTE effect. 

The OH lines may be affected by the UV ``missing opacity'', as discussed by 
Balachandran \& Bell (1998). Allende Prieto \& Lambert (2000) found 
a good agreement between effective temperatures obtained from the infrared 
flux method (IRFM) and from the near-UV continuum for stars with 
$4000\leq T_{\rm eff}\leq 6000$ K when accurate {\it Hipparcos} gravities are 
used. This also agrees with our good reproduction of the near-UV spectral 
region and indicates that the model atmospheres 
used provide an adequate description of the near-UV continuum-forming
region. In any case, even if a poorly understood opacity problem existed as 
described by Balachandran \& Bell, it would have a minor effect on the
OH results since most of the stars in the samples of Israelian et al. (1998),
and Boesgaard et al. (1999a), are hotter than the Sun and very metal-poor. 
It must be mentioned that the OH lines employed by Israelian et al. (1998) and
Boesgaard et al. (1999a) to study the most metal-poor stars in their samples are 
from the (0,0) band, where the corrections to theoretical $gf$ values
are not large. Considering the list used by Israelian et al. (1998), we note that 
only for 3 lines out of 12 these corrections exceed 0.15 dex. 
The same is true for the list employed by Boesgaard et al. (1999a),
who used  $gf$ values from Nissen et al. (1994). The corrections applied
to their $gf$ values are only 0.15 dex. Thus, corrections to oxygen abundances 
due to opacity problems should not exceed 0.15 dex for the stars used in the
present study. This will not change the [O/Fe] vs. [Fe/H] trend significantly.

The forbidden line of oxygen is sensitive to the stellar gravity. 
Israelian et al. (1998) synthesized this line for four 
dwarfs  adopting the same set of stellar parameters as in the OH analysis, 
the $gf$ value given by Lambert (1978), and the equivalent widths provided in 
the literature for the $\lambda$ 6300 \AA\ line. The estimated abundances
were in good agreement with those derived from OH when gravities from 
{\it Hipparcos} are used. This strongly  suggests that a reliable gravity 
scale may indeed be the key to explaining the discrepancies in oxygen abundances
from forbidden and permitted lines in unevolved metal-poor stars. This conclusion
is supported by Mishenina et al. (2000), who also found good agreement between
[O\,{\sc i}] and the triplet for three unevolved halo stars in their sample.
 Nevertheless, all stars with measured [O\,{\sc i}] in Mishenina et al. 
(2000) and Israelian et al. (1998) have [Fe/H]$>$-1.4 and leave open the  
possibility  that differences between oxygen indicators may appear at 
much lower metallicities. Indeed,  FK claimed that there was inconsitent 
results for the two metal-poor stars  BD $+$23\arcdeg 3130 and 
BD $+$37\arcdeg 1458 using stellar parameters derived from an LTE analysis 
of iron lines. These objects were also considered by Israelian et al. (1998) and
Boesgaard et al. (1999a) (only BD $+$37\arcdeg 1458 in the latter case). 
While FK's conclusions for BD $+$37\arcdeg 1458 
do not challenge the linear trend in [O/Fe], their results for 
BD $+$23\arcdeg 3130 are claimed to be inconsistent with those from OH lines.  
However, we argue in the next section  that this discrepancy cannot be 
sustained when a critical analysis of the uncertainties involved in the
determination  from the forbidden line is performed.

It is often claimed in the literature that the abundances derived from the 
forbidden lines are more reliable than those from the triplet and OH. The authors of 
this statement  refer mainly to the fact that the forbidden lines at 6300 and 6363 \AA\ 
are not affected by NLTE effects and are less sensitive to the uncertainties in 
$T_{\rm eff}$. However, results based on the forbidden lines should be interpreted with
caution. Figure 4 demonstrates that very small errors in the equivalent widths (EW) 
of the forbidden line at 6300 \AA\ may
have a dramatic effect on the derived oxygen abundances. Very high signal-to-noise (S/N)
ratio is required in order to minimize this uncertainty; condition which is
normally  not satisfied in most of the studies based on the forbidden line. For
example, Barbuy (1988) used data with a signal-to-noise ratio of 90 and
$R \sim$40.000 in order to derive the oxygen abundance from the forbidden line 
(EW = 4 m\AA) in the most metal-poor giant in her sample: BD $-$18\arcdeg5550 
([Fe/H] = $-$3.). Note that a S/N$\sim$90 optimistically implies about 
$\pm$2 m\AA\ of uncertainty (at 1-$\sigma$ level) in the EW according to
Cayrel's formulae (Cayrel 1988), which introduces an 
uncertainty in the derived abundance of oxygen larger than +0.18 and $-$0.3 dex (Figure 4). 
To this error (which comes from measuring only one weak line) we must add the systematic 
errors due to the uncertainties in gravity (note that Barbuy 1988 used only 3 
lines of each Sc\,{\sc i} and Sc\,{\sc ii} to estimate the gravity) and 
effective temperature. This makes the total error from the [O\,{\sc i}] line large.

Another problem is that the 
gravities of metal-poor giants derived from the LTE Fe analysis are most 
probably underestimated because of the neglect of NLTE effects. This effect 
is already proven to exist in metal-poor dwarfs (Th\'evenin \& Idiart 1999) 
and also in a subgiant (BD +23\arcdeg1330, this study). It has been argued 
(Th\'evenin \& Idiart 1999; Gratton et al. 1996) that 
this effect also operates in metal-poor giants. Work is in progress to study 
NLTE effects on Fe in several metal-poor giants and our preliminary results 
indicate that corrections to gravities derived from the LTE analysis of
Fe can indeed be as large as +0.5 dex. We also note that recently Takeda et al. (2000)
proposed that the [O\,{\sc i}] lines in metal-poor giants may suffer some filling in 
emission that leads to the weakening of the absorption lines. This possibility
has been noticed by Langer (1991), who proposed that the emission from the 
surrounding nebulosity,  formed from the mass ejected in the post-RGB phase, may
account for the observed weakness of the forbidden lines. The fact that some giants
such as HD\,115444 show [O/Fe] about 0.3 dex larger (Westin et al. 2000) than 
``expected'' from  the $plateau$ ([O/Fe] $\sim$  0.4) supports this idea. This is 
also consistent with the fact that there is no discrepancy between [O\,{\sc i}], 
O\,{\sc i}, and OH in several unevolved dwarfs.

\section{The particular case of BD+23\arcdeg 3130}

The analysis carried out by FK is based on gravities derived 
from LTE Fe ionization balance of these subgiants where the NLTE effects 
are strong. In general, NLTE effects play a dominant role in very metal-poor 
stars due to a decrease in electron density. Collisions with free electrons
no longer dominate the kinetic equilibrium as they do in the stars of solar 
composition, and this leads to significant deviations from LTE.
Another consequence of the metal deficiency is an appreciable weakening
in UV blanketing. Both effects lead to a large depletion of 
the \FeI\ atoms affecting the equivalent widths and central depths of
spectral lines (Shchukina \& Trujillo Bueno 2000). In a recent paper,  
Allende Prieto et al. (1999) have 
shown that gravities derived using LTE analyses of iron in metal-poor stars 
do not agree with the gravities inferred from accurate {\it Hipparcos} 
parallaxes, which casts a shadow over oxygen abundance analyses of very 
metal-poor stars based on gravities derived from the LTE Fe ionization 
balance. They find that gravities are systematically underestimated
when derived from ionization balances, and that upward corrections of 0.5 dex or
even higher can be required at metallicities similar to those of our stars. We
remark here that any  underestimation of gravities will also strongly
underestimate the abundances inferred from the forbidden and triplet lines. The
abundances derived from OH will be affected as well but in the oposite
direction.
Th\'evenin \& Idiart (1999) derived gravity corrections of
up to 0.5 dex with respect to LTE values, for the case of stars with 
[Fe/H] $\sim -3.0$. They  have shown that NLTE effects are important
in determining stellar parameters from iron ionization balance. 
 
It has been established (see Rutten 1988, and references therein) 
that a strong overionization of neutral iron in metal-poor stars leads 
to the systematic difference in abundances determined from the \FeI\ and 
\FeII\ lines. This difference increases with decreasing metallicity.
For the \FeII\ lines NLTE effects are found to be unimportant. 
The NLTE modeling predicts  a dependence of the NLTE abundance corrections 
of \FeI\ lines on the lower excitation potential ($\chi$). 
The corrections are particularly large for the low-excitation \FeI\ lines.
Empirical determinations by Ruland et al. (1980), Magain (1988) and
Magain \& Zhao (1996) support this conclusion.
The NLTE abundance corrections in the Sun are in the range 
0.02--0.1~dex, while in the main-sequence stars later than type A  
they can reach 0.3--0.4~dex.

Taking the above into account, we evaluate  the  extent
to which NLTE effects could influence the determination of
the stellar parameters of $\rm BD +23\degree 3130$.

\subsection{NLTE analysis}
\subsubsection{The models and the NLTE solution}

Our study is based on horizontally homogeneous models of LTE atmospheres 
in hydrostatic equilibrium. The grid of interpolated ATLAS9 models is used to
perform NLTE computations spanning the following range of  atmospheric 
parameters: $4800 \leq T_{\rm eff} \leq $ 5130 K, 2 $\leq \log g \leq 3$, and 
$-3<$ [Fe/H] $<-$2.5.
Our NLTE stellar parameter analysis is based on the usual assumptions:
i) iron abundances used to fit EWs of \FeI\ 
        and \FeII\ lines to observed ones have to be independent of 
        excitation potential of the lower level $\chi$;
 ii) the iron abundances have to be independent on the line strengths;
        namely, the stronger \FeI\ lines have on average to give 
        the same abundance as the weaker ones;
iii) the average abundances obtained from \FeI\ and \FeII\ lines have
        to be equal.

The theoretical EWs of  the \FeI\ and \FeII\  lines were
computed using the formal solution of the radiative transfer equation.
NLTE departure coefficients were found from the self-consistent 
solution of the kinetic and radiative-transfer equations using a new
NLTE code, NATAJA (see for details \cite{Shchukina+Trujillo1998};
Shchukina \& Trujillo Bueno 2000).
Here it is enough to mention that we employed novel and effective iterative 
multilevel transfer methods and formal solvers that are capable 
of handling hundreds of transitions efficiently and make feasible 2D and 3D 
multilevel modeling (Auer, Fabiani Bendicho, \& Trujillo Bueno 1994, 
\cite{Trujillo+Fabiani1995}).
Our \FeI\ $+$ \FeII\ atomic model takes into account a multiplet structure
and includes over 250 levels and 500 UV, optical, and IR transitions 
including the regime near \FeI\ continuum. 
The \FeI\ term diagram is, in fact, complete up to $\chi \approx 5.7$~eV.
At higher energies it contains about 50\% of the terms identified 
by Nave et al. (1994).
The \FeII\ diagram is very similar to the atomic model of Gigas (1986).
It contains only the lowest terms with  $\chi < 5.9$~eV
and 25 strong transitions (see Shchukina \& Trujillo Bueno 1998, 2000).

\subsubsection{Spectral lines selected for the analysis}

Notwithstanding such a comprehensive atomic model it contains only 
21  lines from the list of \FeI\ lines 
observed and measured in the spectrum of $\rm BD +23\degree 3130$ by FK. 
All of them except one line cluster into a narrow $\chi$ range between 
2 and 3 eV. We  marked them in  Figures 5 and 6 by large unfilled circles. The only 
line that is outside the range ($\lambda 7511.02$~\AA ) has  $\chi$ =4.16 eV.

This selection effect prevents us from using the \FeI\ abundance versus 
$\chi$-trend as a criterion of correctness of the stellar parameters. To avoid 
this shortcoming we tried to extend our line list to larger $\chi$-samples. 
According to FK, LTE synthesis of the 58 \FeI\ lines for the atmospheric
model with parameters derived by these authors  ($T_{\rm eff} = 4850$, 
$\log g$ = 2, and $\log \epsilon=4.69$) produces very good agreement
with observations (the standard deviation, $\sigma$, is only 0.04 dex). 
If we assume that these stellar parameters reproduce both the EWs 
from FK, and that such  modeling of \FeI\ lines is  correct not only 
for their observed list  of  \FeI\ lines but also  for other \FeI\ lines located
in the same wavelength region (4000--8000 \AA),  then we can consider 
synthetic LTE equivalent widths of the latter lines as ``observed'' ones.  
Following this idea we have selected  from our atomic model 98 \FeI\ lines 
and calculated their LTE equivalent widths using the model atmosphere
with stellar parameters obtained by FK. Furthermore, 
these equivalent  widths, ranging from 1.5 to 80 m\AA, are consider as
``quasi-observations''. The wavelengths of these \FeI\ lines, their
$\chi$-values, and oscillator strengths are listed in Table 5.
Note that these lines are distributed rather uniformly with $\chi$.
In addition to the line cluster between 2 and 3 eV 
the new  set contains 13 \FeI\ lines with $\chi <$ 2~eV,
27 lines in the range 3--3.8 eV and 17 lines with $\chi >$ 3.8~eV.
Using our list of  Fe lines we obtain with the
stellar  parameters of FK the same LTE Fe abundance claimed in their paper.
It is interesting to note that everywhere in our figures, abundances 
for the aforementioned high-excitation \FeI\ line $\lambda 7511.02$~\AA\ 
deviate considerably in comparison with those for other high-excitation lines.
For \FeII\ lines we used the {\it real} observations obtained by FK,
which includes seven  lines.

\subsection{NLTE abundance correction}

Our NLTE simulation reveals that  \FeII\ lines
in all the atmospheric models considered
do not suffer in fact from NLTE effects. 
However, this is not the case for the  \FeI\ lines.
The LTE  abundances derived from the \FeI\ lines are lower compared with 
NLTE values. Our computations clearly demonstrate that for the fixed value of 
metallicity the NLTE corrections are very sensitive to the changes in 
$T_{\rm eff}$ and $\log g$. The NLTE corection increases with $T_{\rm eff}$ due to the 
ultraviolet overionization effects. For the same $T_{\rm eff}$  a larger gravity 
results in larger NLTE abundance corrections. The mean $\delta \log \epsilon$ 
are in the range 0.2 -- 0.35~dex.

The NLTE abundance corrections for individual \FeI\ lines
and for the  stellar parameters of $\rm BD +23\degree 3130$ suggested by FK 
are plotted in Figure 5.
The  abundance corrections  are largely independent
on the excitation potential up to $\chi \approx 2$~eV.
However, $\delta log \epsilon$  decreases when the lines with 
$\chi > 2$~eV are considered.
The mean value of the NLTE abundance correction  $\delta log \epsilon$  
for the \FeI\ lines is  0.22~dex while for the \FeII\ lines there are no 
corrections for departures from LTE.
Although the standard deviation of the abundance derived from \FeI\ lines 
is not very large (0.07~dex) there is a trend  with $\chi$ and 
a clear gap of 0.11~dex between the abundances derived from the ``low''
($\chi < 3.8$~eV) and ``high'' ($\chi > 3.8$~eV) excitation lines.
In general, the gap is smaller in the case of LTE. It is also interesting 
to point out that dependence on the line strength is less pronounced.

Our computations confirm the results by Thevenin \& Idiart that the LTE Fe analysis 
in very metal-poor stars strongly underestimates gravities and metallicities.
If we take into account the NLTE abundance corrections, the stellar parameters
suggested by FK do not meet the condition requiring equal abundances derived
from the \FeI\ and \FeII\ lines. The abundance difference due to neglect
of the NLTE effects in the ionization equilibrium balance amounts to 
0.2~dex. We found that the parameters used by Israelian et al. (1998) 
do not meet the conditions listed in 5.1.1 either.
Figure~6 shows that a model similar to the one used by FK but with a gravity 
which is higher by 0.5~dex, is able to improve the situation considerably.
With such model the difference in abundances derived from \FeI\ and \FeII\ lines 
becomes only 0.026~dex while the standard deviation from the mean abundance 
derived from the \FeI\ lines is 0.06. The gap between ``low''
($\chi < 3.8$~eV) and ``high'' ($\chi > 3.8$~eV) excitation lines 
has also decreased to 0.07~dex.

\subsection{Uncertainties in the NLTE analysis}
Classical abundance determination depends on  a large number of parameters.
We evaluated the sensitivity of the abundances, $\log \epsilon$, derived from
the  \FeI\ and \FeII\ lines
to the uncertainties of  the most important parameters.

Several tests showed that a change in the enhancement factor, $E$, of 
the van der Waals damping constant,
$\gamma_6$,  from 1.5 to 2.0 produces changes of less than 
0.04 dex in the \FeI\ abundance and can be neglected.
The strongest low-excitation lines display the largest errors.
The lines with equivalent widths EW $ < 80$~m\AA\ are
insensitive to the uncertainties in $E$-factors. 
Fortunately,  all  \FeI\ and \FeII\ lines from Table~5 
have equivalent widths  smaller than this value.

The values of microturbulent velocity (V$_{t}$) used by Israelian et al. (1998), 
and FK for BD$ +23\degree 3130$ differ in 0.35 \kms. 
Our tests show that \FeI\ lines with EW $ \approx 80$~m\AA\ 
and $\chi <3$~eV are most sensitive to V$_{t}$.
However, even in that case the influence of the 
microturbulence uncertainties could be reduced or even eliminated 
by using lines with EW $ < 30$~m\AA. It turned out that the  change 
in $\log \epsilon$ corresponding to 0.35 \kms\ for the \FeI\ lines
listed in Table~6 is  $\sim 0.04$~dex.
Thus, the influence of uncertainties in the microturbulent velocity 
on the abundance derived from \FeI\ lines is negligibly small.

The accuracy of the oscillator strengths
has a strong influence on the dispersion and the mean value 
of the iron abundance derived from \FeI\ and \FeII\ lines.
In our study we used for low-excitation lines 
($\chi <3$~eV) mainly  the Oxford $gf$ values
(\cite{Blackwell+Ibbetson+Petford+Shallis1979},
\nocite{Blackwell+Petford+Shallis1979} 1979b,
\nocite{Blackwell+Petford+Shallis+Simmons1980} 1980,
\nocite{Blackwell+Petford+Shallis+Simmons1982} 1982a,
\nocite{Blackwell+Petford+Simmons1982} 1982b,
\nocite{Blackwell+Shallis+Simmons1982} 1982c,
\nocite{Blackwell+Booth+Petford1984} 1984,
\nocite{Blackwell+Booth+Haddock+Petford+Leggett1986} 1986,
\nocite{Blackwell+LynasGray+Smith1995} 1995).
The Oxford set does not contain any lines with $\chi >3$~eV.
Our sample of the high-excitation lines is dominated by $gf$ values
measured by
O'Brian et al. (1991) and the
Kiel--Hannover group
      (Bard, Kock, \& Kock 1991;
      \cite{Holweger+Bard+Kock+Kock1991},
       \nocite{Holweger+Kock+Bard1995} 1995).
There is also a large set of the \FeI\ lines whose measured 
oscillator strengths are taken from the old sources (compilation by 
\cite{Fuhr+Martin+Wiese1988}). 

Several authors (\cite{Grevesse+Noels1993};
\cite{Lambert+Heath+Lemke+Drake1996};
\cite{Grevesse+Sauval1998})  suggest that
$gf$ values of the higher-excitation \FeI\  lines
that come from laboratory measurements provide likely less accurate
results than those from the $gf$ values of low-excitation lines.
The use of the oscillator strengths mentioned above
opened  a new ``iron problem'' (\cite{Grevesse+Noels1993})
`namely the behavior of the abundance derived from \FeI\ lines \vs\  
the excitation energy for a very large range of $\chi$, from 0 to 7.5 eV, 
observed in the solar spectrum'.
The solar photospheric abundances found from the low-excitation lines using 
$gf$ values of the Oxford group turns out to be systematically larger than
those from high-excitation lines obtained with $gf$ 
measured by Kiel--Hannover group and by O'Brian et al. (1991).
Since $gf$ values and abundance values enter the line extinction
coefficient as a product,
such  behavior might be attributed to a $gf$ systematic trend of about
0.06 dex
(\cite{Bard+Kock+Kock1991};
\cite{Holweger+Kock+Bard1995};
\cite{Lambert+Heath+Lemke+Drake1996}).
Another point of view is to attribute the trend  
to differences in the atmospheric properties of low- and 
high-excitation lines
(\cite{Blackwell+LynasGray+Smith1995})
or both effects together
(Kostik, Shchukina, \& Rutten 1996;
\cite{Grevesse+Sauval1998}).

The most accurate \FeII\ oscillator strengths 
are not yet as accurate as for the \FeI\ lines. Following
FK we used $gf$ values obtained by Bi{\'e}mont et al. (1991)
and Kroll \& Kock (1987). A discussion of these and other sources of
the oscillator strengths has been given recently by 
Grevesse \& Sauval (1999) and Lambert et al. (1996). 
The mean uncertainty in the $gf$ values for  \FeII\ lines
is found to be in the range $\pm$ 0.05--0.07 dex, which corresponds
to an abundance dispersion $\sim 0.1$~dex.
Uncertainties of the NLTE abundance corrections $\delta \log \epsilon$ 
are discussed in detail by Shchukina \& Trujillo Bueno (2000).
The values $\delta \log \epsilon$ are not very sensitive to  
uncertainties in collisional rates, photoionization cross-sections,
continuum opacity, etc. The errors introduced by uncertainties in 
these parameters may lead to  errors in the  mean NLTE abundance, 
$\delta \log \epsilon$, of about $\pm 0.03$~dex.

\subsection{Errors of the stellar parameter determinations}

Summing up  the analyses completed above, we conclude
that the largest errors in the iron abundance determination using   
\FeI\ lines  are expected to be caused by neglecting the NLTE effects. 
The largest errors in the abundance derived from  \FeII\ lines 
are caused by the uncertainties in the oscillator strengths
(if we exclude the errors in EW measurements).
To what extent can these errors influence the stellar parameters
determination for the star $\rm BD +23\degree 3130$?

Our computations show that the variations in $T_{\rm eff}$ primarily affect 
 the  abundance derived from \FeI\ lines
but make little difference to the abundance from  \FeII\ lines.
Abundance changes for  \FeI\ lines that result from a 270 K
change in effective temperature are in the range
$\sim 0.26$ for LTE case  and $\sim 0.36$  dex for NLTE. 
In the case of \FeII\ lines  they are an order of magnitude smaller.
On the contrary, for  \FeII\ lines the gravity dependence is much more
pronounced, while for \FeI\ lines it is  negligibly small. 
As a result, small errors in the \FeI\ abundances can only be compensated
for by large changes in the assumed gravity. Or small errors in \FeII\ 
abundances can only be fixed by large changes in the assumed 
effective temperature.
Our computations show that for the NLTE $\delta \log \epsilon$ 
range 0.2--0.35~dex  obtained using \FeI\ lines the correction in 
$\log g$ may reach $\sim$ 0.5~dex in the cooler model ($T_{\rm eff}=4850$ K)
or even more in the hotter one ($T_{\rm eff}=5130$ K).
An abundance dispersion of 0.1~dex for \FeII\ lines causes dispersion
of $T_{\rm eff}$ of 100 K. 
To reduce the latter effect it would be extremely desirable 
to enlarge the set of \FeII\ lines.
It is obvious that the small seven-line sample can be a source 
of serious troubles in  stellar parameter determinations.

\subsection{The ``best-choice'' parameters of  $\rm BD +23\degree3130$ and 
its oxygen abundance}

Our computations are aimed at finding the NLTE solution that  gives the 
minimum standard deviation, $\sigma$, of the mean abundance
and satisfies the criteria discussed in Section 5.1.1.
If we neglect the offset in abundances derived from the
high-excitation ($\chi > 3.8$~eV) and low-excitation ($\chi < 3.8$~eV)
lines, then the best model is the one displayed in Figure 6. 
The ``best-choice'' stellar parameters for this NLTE case turn out to be  
$T_{\rm eff} = 4825$ K, $\log g = 2.5$, and [Fe/H] = $-$2.66. 
The  top panel shows 
the NLTE equivalent widths  for the atmospheric model represented by these 
parameters against LTE values computed for the model with parameters by 
FK. Agreement between ``quasi-observed'' equivalent widths 
and those obtained in NLTE is quite satisfactory. However,
if the \FeI\ line sample is divided into a ``low'' ($\chi < 3.8$)~eV 
and ``high'' ($\chi > 3.8$)~eV samples, the mean NLTE abundances are 
4.86 $\pm$ 0.05 and 4.79 $\pm$ 0.06 for the ``low'' and ``high'' samples, 
respectively. This means that even this solution cannot be considered
as $final$ given the abundance offset found for the two samples of Fe lines.

We could not find a model which did not show an abundance gap between 
``low-'' and ``high-'' excitation lines. This failure can be attributed to
our poor knowledge of the atmospheric model structures and is possible due 
to some 3D and/or hydrodynamical effects.

This shortcoming can possibly be avoided if we concentrate our study 
only on the high-excitation lines with $\chi > 3.8$ eV. This approach
can be justified for the following reasons: a) the high-excitation lines
are sensitive to NLTE effects in much less degree than low excitation ones
and therefore to many uncertain
factors such as overionization, granulation, geometry, etc.;  b)
the high-excitation lines are insensitive to the microturbulence; 
c) they are also  insensitive to the damping factors;  
d) there probably exists a $\log gf$ versus $\chi$  trend for
the low excitation lines; and e) the high excitation lines
are less affected by 3D and hydrodynamical effects (Shchukina \& Trujillo Bueno 2000,
Grevesse \& Sauval , 1998, 1999).

Limiting our computations to 17 high-excitation lines, we
are left with only one criterion for the selection of the ``best model''.
The only demand on the model parameters will be the minimum abundance scatter
derived from the high-excitation Fe lines and the agreement between log $\epsilon$(Fe\,{\sc i})
and  log $\epsilon$(Fe\,{\sc ii}). Results from four models are presented in Table 6. 
It is clear that the model with $T_{\rm eff}$ = 5130 K and $\log g$ = 3.0 is the ``best
choice''. We can give more weight to this model since the gravity derived
from the {\it Hipparcos} parallax is $\log g$ = 3.05 $\pm$ 0.25.  

An important conclusion that follows from the NLTE analysis of \FeI\ lines
is that the surface gravity of the star  $\rm BD +23\degree 3130$
derived using LTE approach is in error by at least $\sim 0.5$~dex. In addition, 
the metallicity of the star has to be increased as well using the mean NLTE
abundance correction, $\delta \log \epsilon \approx 0.2$~dex.

\subsubsection{The oxygen abundance}

FK did not detect the [O\,{\sc i}] line at 6300 \AA\ establishing an upper 
limit to its equivalent width of 1 m\AA. They argued that the parameters used by 
Israelian et al. (1998) would predict an [O\,{\sc i}] absorption at 6300 \AA\ 
with a larger EW which was not observed. However, given the S/N$\sim$ 250 of 
their spectrum of BD $+$23\arcdeg 3130, one cannot claim a detection of absorption 
features with EW $\leq$ 2 m\AA, and errors in the EW for any barely detected 
line would have been at least 0.7 m\AA\ (although a more realistic
value would be 2 m\AA),  implying  +0.2 and $-$0.5 dex error in the 
[O/H] value (see Figure 4). Moreover, to this error we must add the systematic 
one due to uncertainties in gravity and  effective temperature, which makes the 
total error in the oxygen abundance from the [O\,{\sc i}] line very large.  
In conclusion, the failure to detect the forbidden  line cannot be used to 
question the validity of the abundances derived from  the OH lines and triplet.
The new really  high S/N measurement of the forbidden line in this
star performed by Cayrel et al. (2000), using UVES at VLT, of EW=1.5$\pm$0.3 
m\AA\ is perfectly consistent with our determinations of OH and triplet lines 
definitely solving the controversy. We show in what follows that there 
is consistency between the abundances inferred from the three oxygen indicators.

Assuming our stellar parameters, the EW measurement by Cayrel et al. (2000)  
for the forbidden line, our new triplet observations  and the  OH measurements 
by Israelian et al. (1998),  we find  
[O/H]=$-$1.61$\pm$0.15, $-$1.5$\pm$0.14, $-$1.45$\pm$0.33, respectively. There is
a clear agreement between these three indicators given the error bars from 
each measurement. We interpret this as an indication that 
any 3D correction to the OH abundance,  or NLTE correction for triplet 
abundance, is on the order of  0.15 dex. The agreement is even better 
(within 0.07 dex) if we employ models without convective overshooting. We conclude  
that in this cool subgiant it is possible to find a consistent determination 
of oxygen abundances using 1D models and  stellar 
effective temperature from IRFM based calibration and  gravity from 
{\it Hipparcos} parallax and/or Fe NLTE analysis. The plot in figure 7 
shows the difference in oxygen abundances derived for OH and [O\,{\sc i}]
in a number of stars whose parameters have been estimated in this way. These
stars cover the metallicity range from $-$1 to $-$2.5 and strongly support the
reliability of the linear trend in oxygen abundances as metallicity decreases.
Four of these stars are listed in Table 4 of Israelian et al. (1998) and two
others were studied by FK. The [O\,{\sc i}] abundance ([O/H]=$-$1.52) for the star BD$+$37\arcdeg 1458 
([Fe/H]=$-$2.1) has been estimated from the equivalent width of the [O\,{\sc i}] 
line using an {\it Hipparcos}-based gravity ($\log g$=3.35$\pm$0.21) and 
effective IRFM temperature $T_{\rm eff}=5260\pm100$ K from Israelian et al. (1998).
Using the same stellar parameters we find excellent agreement with the oxygen abundance
[O/H]=$-$1.60 derived from the OH lines observed by Israelian et al. (1998).

Using a different calibration, Carretta et al. (2000) have claimed that their 
oxygen NLTE analysis yields the same O abundances from both permitted and forbidden 
lines for stars with $T_{\rm eff}> $ 4600 K. If this had been the case,
there would have been no problems with oxygen abundance derived from 
the triplet and the forbidden line in BD +23\arcdeg 3130.
Assuming the parameters of FK for this star, one can find from the triplet an 
LTE abundance [O/H] = $-$1.39. In this parameter range the NLTE abundance correction
for the triplet is about $-$0.1 according to Carretta et al. (2000) and
Mishenina et al. (2000). Thus, the [O/H] = $-$1.39 ratio
obtained from the triplet is in odds with [O/H] = $-$2.49 limit claimed from the 
undetected [O\,{\sc i}] line by FK.

\section{Discussion}

\subsection{A {\it plateau} versus {\it linear trend}}

Our results support a linear increase of [O/Fe] with decreasing metallicity 
(Table 7 and Figures 8,9,10 and 11). This is in agreement with some previous analyses 
(Abia \& Rebolo 1989,  Tomkin et al. 1992, King \& Boesgaard 1995, 
Cavallo et al. 1997, Israelian et al. 1998,
Boesgaard et al. 1999a, Mishenina et al. 2000, Takeda et al. 2000) but
at odds with others (Barbuy 1988, Kraft et al. 1992, Carretta et al. 2000);  
there appears to be a clear dichotomy among authors who support
a plateau in [O/Fe] for stars with [Fe/H]$<-$1 and those who do not.
The situation is made even more confusing by the fact that there is 
an overlap both of objects and of O indicators used by authors arriving
at different conclusions. The comparison we made with the results
of FK and Carretta et al. (2000), suggest that the
discrepancies are rooted in the atmospheric parameters adopted by different
authors. Differences in the observational data, in the 1-D models and NLTE
treatment of the O\,{\sc i} triplet lines  are second order effects.
The set of atmospheric parameters chosen is always debatable, however
we wish to point out that our choice provides gravities which are in
agreement with {\it Hipparcos} parallaxes and achieve consistency among
different O indicators, when available. We independently derived 
metallicities from our spectra and only in one case from EWs taken 
from the literature (for G271-162). We discard any significant change 
in the slope of the [O/Fe] versus [Fe/H] relation in the whole metallicity 
range from  0 to $-$3.1. The corresponding increase in Fig. 11 can be fitted
using a straight line with a slope $-$0.33$\pm$0.02 (taking the
error bars into account).

The existence of a $plateau$ with [O/Fe] $\sim$ 0.4 at [Fe/H] $< -1$ in metal-poor 
giants could be questioned. The results of Takeda et al. (2000) show 
that metal-poor giants may have [O/Fe] $>$ 0.6 in the metallicity range 
[Fe/H] $< -2$. To some extent different conclusions on the existence of 
a plateau may be derived from the same data set. Tomkin et al. (1992)
point out that ``.. a linear least sqares fit to the 10
field giants ($-3 \leq$ [Fe/H] $\leq-$1.7) in Fig. 8 of Sneden et al
provides a slope of $-0.33\pm 0.10$ for [O/Fe] vs [Fe/H]...''.
It is interesting that this slope derived from the data of Sneden
et al. (1991) is in excellent agreement with our value and the slopes found
by Israelian et al. (1998) and Boesgaard et al. (1999a), who provide
$-0.31\pm +0.11$ and $-0.35 \pm 0.03$, respectively. In spite
of this, the results of Sneden et al. (1991) are usually quoted as evidence
against the plateau probably because they obtained [O/Fe] ratios 
lower than 0.6. 

So what is the [O/Fe] vs. [Fe/H] relation for giants ? We can 
think of two possibilities. There is a real scatter
in the range 0.3 $<$ [O/Fe] $<$ 1 at [Fe/H] $< -1$, which could  be explained
by a weakening of the forbidden lines in some of the evolved giants due to the presence
of  circumstellar matter. Indeed, differences in the published EWs of the 
forbidden line for  some giants may support this hypothesis (work
in preparation). If this is the case, then one cannot display both giants and
unevolved subdwarfs on the same [O/Fe] vs. [Fe/H] graph (e.g., Carretta et al. 2000). 
It is also possible that giants follow the linear trend just as unevolved subdwarfs,
but previous investigators have failed to discover this for two main reasons; 1) they
did not use sufficiently high S/N (e.g., S/N $>$ 400) to investigate uncertainties 
involved in the determination of oxygen abundance from the forbidden line at 6300 \AA\
in the most metal-poor targets of their samples and 2) the stellar parameters of the
most metal-poor giants were not correct because  NLTE effects were not taken 
into account.  However, we cannot exclude the possibility that the plateau
derived from the [O\,{\sc i}] in giants is correct.
Work is in progress to study the stellar parameters of several giants
with [Fe/H] $< -2$ by performing a NLTE analysis of Fe lines and investigating
its impact on the oxygen abundances derived from OH, forbidden, and triplet
lines.

\subsection{Oxygen nucleosynthesis in the early Galaxy}

The new data in ultra-metal-poor unevolved stars confirm  the previous
findings for a progressive  rise in oxygen overabundances as we go to
[Fe/H]$ < -$2.5. In particular, the high [O/Fe] ratio in G64-12 provides
evidence for enhanced oxygen production in the first nucleosynthesis events in 
our Galaxy. Can these high [O/Fe] ratios in ultra-metal-poor stars be 
understood in terms of massive star nucleosynthesis models? 

The first attempts to explain the linear trend were made by 
Abia, Canal, \& Isern (1991), and more recently Goswami \& Prantzos (2000) 
have suggested three possible scenarios to explain
the observed linear trend of [O/Fe]. The first possibility, which has 
already been explored in the literature, considers the early evolution
of Type Ia SNe, which started to contribute Fe to the ISM already in 
the epoch when [Fe/H] $\sim-$3.0. Goswami \& Prantzos (2000)
remark that this model cannot be accepted, as it also predicts similar linear
trends for other $\alpha$ elements which are not observed. However, we would like
to recall that the ``traditional'' trend of $\alpha$ elements (i.e.,
a unique $plateau$ at [Fe/H] $< -1$) has been challenged recently by 
Idiart \& Th\'evenin (2000) on the basis of NLTE computations of $\alpha$ 
elements. In addition, the observations indicate 
(Francois 1987, 1988) that the [S/Fe] ratio increases from approximately 0 to 0.7 
in the metallicity range $-1.5 <$ [Fe/H] $<0 $. There is no $plateau$ observed at
[Fe/H ]$< -1$, and the models fail to account for the observed [S/Fe] ratios 
(Goswami \& Prantzos 2000). This is confirmed by recent observations of Takeda et al.
(2000), who report [S/Fe] = 1.11 in the metal-poor ([Fe/H ]= $-$2.72) giant HD\,88609.
As a second possibility, Goswami \& Prantzos (2000) mention the
 metallicity-dependent oxygen yields 
proposed by Maeder (1992). Given that the stellar mass loss
at very low metallicity is poorly understood, one can propose a model in which
[O/Fe] increases below [Fe/H] $\sim -1$, while [$\alpha$/Fe] is constant. However, 
Goswami \& Prantzos (2000) argue that even this model is not acceptable,
as it predicts [C/Fe] and [N/Fe] ratios increasing with [Fe/H] (which are not 
observed). The third model proposed by Goswami \& Prantzos (2000) considers the possibility 
of having metallicity-dependent Fe and $\alpha$-element yields at [Fe/H] $< -1$. 
This is possible if/when the supernova layers inside the 
carbon-exhausted core are well mixed during the explosion (to have [$\alpha$/Fe] 
$\sim$ const., in the ejecta for any [Fe/H]). This last model requires that the oxygen, 
carbon, and helium layers will escape with the same yields, independent on the 
metallicity, as they are loosely bound. Note that these models
did not take into account stellar rotation which plays a very important role in 
low-metallicity massive stars (Maeder \& Maynet 2000). Strong observational 
support for  matter mixing, which
occurs in  SN ejecta during or prior to explosion comes from the analysis 
of the low-mass X-ray binary system Nova Sco 1994 (Israelian et al. 1999). 
If the mass of the black hole in Nova Sco 1994 is more than $ \sim$ 4
$M_{\sun}$ (which is just a lower limit given by observations),
then it must have ``eaten'' all Fe-group elements plus the $\alpha$ elements
Ti, Ca, and S. Some amount of Mg and Si, together with almost all the O,
may escape the collapse and therefore appear in the supernova ejecta captured by 
the secondary. However, observations show that all $\alpha$ elements are uniformly 
enhanced in the atmosphere of the secondary star, and that therefore there
must have been some mixing in the supernova ejecta. Most probably, this system is 
a relic of a hypernova that left a massive black hole and ejected similar amounts 
of $\alpha$ elements (Israelian et al. 1999; Brown et al. 2000).

The yields from standard Type II supernovae have been investigated by a number of authors 
(e.g., Woosley \& Weaver 1995; Thielemann, Nomoto, \& Hashimoto 1996). It is well known
that stellar yields are influenced by metallicity. Low mass-loss rates in very 
metal-poor massive stars lead to more massive He cores, and therefore more He
is converted into oxygen (Maeder 1992). In contrast, the most massive 
solar-metallicity stars ($M_{\rm ZAMS} > 40\,M_{\sun}$) go through a
Wolf--Rayet phase where the mass loss is very efficient
and end up with low-mass (4--6 $M_{\sun}$) He cores (Maeder \& Meynet 1989). 
It is well known (Woosley 1996; Fryer, Woosley \& Heger 2000) that 
during the contraction phase of helium cores 
with $M \geq 40~M_{\sun}$ the temperature in the center of the star gets 
very high ($\geq~3\times10^9$K) while the density remains low.  This favours
the electron-positron pair instability which leads to explosive oxygen
or even silicon burning.
 It is interesting that the pair-creation supernovae produced by stars with 
$M_{\rm ZAMS} = $ 150--200$M_{\sun}$ and $Z=0.0004$ lead to the formation 
of CO cores with $M_{\rm CO}$ = 60--100 $M_{\sun}$. A complete 
thermonuclear disruption of these objects produces very large amounts of oxygen and 
carbon (see Figure 2 in Portinari 2000). As a matter of fact, the maximum amount of 
oxygen at $Z=0.0004$ is produced not by 20--25 $M_{\sun}$ progenitors but by 
$M_{\rm ZAMS}$ = 150--200 $M_{\sun}$ stars (Portinari 2000). 
Thus, it is possible that the early ISM of our Galaxy has been polluted 
by very massive CO cores of the first pair-creation supernovae. This
idea is supported by Qian \& Wasserburg (2000), who on the basis of three-component
mixing model for the evolution of O relative to Fe, suggest a linear rise
of [O/Fe] if the contribution of the first very massive stars is taken
into account. The stability of ultra-metal-poor very massive stars was 
investigated recently by Baraffe, Heger \& Woosley (2000) who concluded that
such stars could take an active role in the nucleosynthesis of the early Galaxy.

Our observations suggest that the Fe production sites were active already in 
the early halo. Recently, Nomoto et al. (1999) have presented a model
of an exploding CO core with a mass of 12--15 $M_{\sun}$, and an explosion
energy of 2--5 $\times$ 10$^{52}$ erg in order to explain the observations
of the hypernova SN1998bw. The progenitor of the CO core initially had
$M_{\rm ZAMS}=40~M_{\sun}$ and large angular momentum. Placing  a mass cut-off
at 2.9 $M_{\sun}$ Nomoto et al. (1999) have computed that 0.7 $M_{\sun}$ of
$^{56}$Ni is ejected; the amount required to explain observations of SN1998bw. 
It is possible that the linear trend can be explained by an increasing role
of hypernovae in the early epochs of the formation of the Galaxy 
(i.e., [Fe/H] $< -1$). The estimated Galactic rate of Type Ic hypernovae 
is $10^{-3}$~yr$^{-1}$ (Hansen 1999) but it could have been  much higher in the 
early Galaxy. Small mass-loss rates due to the low metallicity in the first
generations of massive halo stars could have led to large helium cores and
the conservation of the primordial angular momentum. Since hypernovae
are favored by stars with a large helium core mass and rapid rotation,
we anticipate significant sulfur (e.g. Takeda et al. 2000) and iron production 
in the early Galaxy following the yields computed by Nomoto et al. (1999).
Whether the objects which started to produce large amounts of Fe in 
the very early halo were metal-poor progenitors of the first hypernovae,  
very massive stars, or other types of supernovae remains to be solved by
future investigations.

Accelerated protons and $\alpha$ particles in cosmic rays interact with 
ambient CNO in the 
ISM and create beryllium and boron. According to the standard 
Galactic cosmic ray (GCR) theory, these interactions in the general ISM 
should have given a quadratic relation between these elements and O that 
is not observed. Alternatively, spallation of cosmic ray CNO nuclei 
accelerated out of freshly processed material could account for the primary 
character of the observed early Galactic evolution of Be and B. Another 
production site is the collective acceleration by SN shocks of ejecta-enriched 
matter in the interiors of superbubbles. In these two cases, the evolution of 
Be should reflect the production of CNO from massive stars.

The dependence of $\log$ (Be/H) on [Fe/H] and on [O/H] is essentially linear 
but with different slopes:  $\sim 1.1$ and $\sim 1.5$, respectively, and  
similar behavior is found for boron (Molaro et al. 1997, 
Garc\'\i a L\'opez et al. 1998; Garc\'\i a L\'opez 1999; Boesgaard et al. 1999b, ). 
Three types of GCR models exist at present that try to explain their observed 
evolution. These are 1) a pure primary GCR from superbubbles (Ramaty et al. 2000), 
2) a hybrid model based on GCR and superbubble accelerated particles 
(Vangioni-Flam \& Cass\'e 2000), which could be accomplished by a pure superbubble 
model (Parizot \& Drury 2000), and 3) standard GCR (Olive 2000, Fields \& Olive 1999). 
The models presented by Ramaty et al. (2000) and Olive (2000) show more consistency 
when a steady increase in [O/Fe] with decreasing metallicity is adopted. 
In addition, Ramaty et al. (2000) have proposed that a delay between the 
effective deposition times into the ISM of Fe and O (only a fraction of 
which condensed in oxide grains) can  explain a linear trend in [O/Fe].
Their model also predicts a linear rise in [S/Fe].

\section{Conclusions}

Oxygen abundances in several unevolved ultra-metal-poor stars have been derived
using UV OH and O\,{\sc i} IR triplet lines. It is found that the  new
abundances from both indicators  confirm  the linear trend of [O/Fe] vs. [Fe/H] first
 reported by Abia and Rebolo (1989) and more recently by Israelian
et al. (1998) and Boesgaard et al. (1999a). In G64-12, the most metal-poor star 
in our sample,  we find the highest  [O/Fe] ratio at 1.17. Our best
estimate of the slope in the trend [O/Fe] versus [Fe/H] is $-0.33\pm$0.02.
 
A detailed NLTE analysis of Fe\,{\sc i} lines has been carried out for the
subgiant star BD+23 3130 providing more reliable stellar parameters, in
particular a surface gravity in good agreement with the value derived from its
accurate {\it Hipparcos\/} parallax. Using these parameters and taking into account 
the uncertainties involved in deriving oxygen abundances from the weak forbidden
line, we argue that the discrepancy noticed by FK can no longer be sustained.
New measurements of the forbidden line by Cayrel et al. (2000) confirm
our conclusions. There is no discrepancy between OH, triplet and forbidden
line in BD+23 3130 when 1D ATLAS9 models are employed. Similar good
agreement between the OH lines and the forbidden lines is found for several 
other metal-poor unevolved stars.

Examination of several scenarios for nucleosynthesis in low-metallicity 
Type II supernova and/or hypernovae provides a variety of possible explanations 
for the increase of oxygen overabundance with decreasing metallicity in the early Galaxy.

\section{Acknowledgments}

The data presented here were obtained with the William 
Herschel Telescope, operated on the island  of La Palma by the Isaac Newton 
Group in the Spanish Observatorio  del Roque de los Muchachos 
of the Instituto de Astrof\'\i sica de  Canarias, the
public data released from the UVES commissioning and Science
Verification at the VLT Kueyen telescope, European Southern Observatory, 
Paranal, Chile, as well as with the  W. M. Keck Observatory,
which is operated as a scientific partnership among the California
Institute of Technology, the University of California, and the
National Aeronautics and Space Administration. The Observatory
was made possible by the generous financial support of the
W. M. Keck foundation. We are grateful to Ya. V. Pavlenko for providing 
the code {\sc wita3} and for helpful discussions. N. S. would like to 
thank R.~Kostik for several useful discussions. We thank Carlos Allende Prieto 
for many discussions about {\it Hipparcos} gravities of the program stars and for
providing us the gravity values for BD +23\arcdeg3130 and BD +37\arcdeg1458 . 
Ilia Ilyin is thanked for helping with the reduction of NOT/SOFIN spectra. This 
research was partially supported by the Spanish DGES under projects PB95-1132-C02-01 
and PB98-0531-C02-02 and also by a NATO grant 950875.


\doublespace

\clearpage

\begin{deluxetable}{lrcccrrrr}
\scriptsize
\tablecaption{Observing log. 
\label{tbl-2}}
\tablewidth{0pt}
\tablehead{
\colhead{Configuration} & 
\colhead{$\lambda$/$\Delta \lambda$} &
\colhead{Spectral region (in \AA)} &
\colhead{Date} & 
\colhead{Target} & 
\colhead{$V$} &
\colhead{Exp. time (s)} &
\colhead{S/N} & 
}
\startdata
WHT/UES      & 30\,000 & 5280--10300 &  1999 Jul 26      & G64-37             & 11.13 & 6000  & 150  \nl
WHT/UES      & 30\,000 & 5280--10300 & 1999 Jul 27    & G268-32            & 12.11 & 7500  & 150  \nl
WHT/UES      & 50\,000 & 4120--8100  & 1997 Jun 15       & G64-37             & 11.13 & 3600  &  60  \nl
WHT/UES      & 50\,000 & 4930--7940  & 1996 Oct 25  & BD $+$23\arcdeg 3130   & 8.95  & 3600  & 250  \nl
WHT/UES      & 50\,000 & 4930--7940  &  1996 Oct 25      & G4-37              & 11.42 & 13\,600  & 180  \nl
WHT/UES      & 50\,000 & 4930--7940  &  1996 Oct 25  & BD $+$03\arcdeg 740    & 9.81  & 3600  & 220  \nl
NOT/SOFIN    & 30\,000 & 3800--10600 & 1999 May 31   & BD $+$23\arcdeg 3130   & 8.95  & 2400  & 50  \nl
NOT/SOFIN    & 30\,000 & 3800--10600 & 1998 Oct 06       & LP 831-70          & 11.62 & 6600  & 80  \nl
NOT/SOFIN    & 30\,000 & 3800--10600 & 1998 Oct 07       & G268-32            & 12.11 & 4800  & 60  \nl 
KeckI/HIRES  & 60\,000 & 6430--8750  & 1999 Dec 4      & G275-4             & 12.12 & 3600  & 170  \nl
KeckI/HIRES  & 60\,000 & 6430--8750  & 1999  Dec 4     & LP 831-70          & 11.62 & 2400  & 160  \nl
KeckI/HIRES  & 60\,000 & 6430--8750  &1999  Dec 4      & G268-32            & 12.11 & 3600  & 170  \nl
KeckI/HIRES  & 60\,000 & 6430--8750  & 1998 Jun 24      & G275-4             & 12.12 & 1200 & 100  \nl
VLT/UVES     & 110\,000 & 6052--8205 &  1999 Oct 11,12,13 & G271-162           & 10.35 &  10800 & 300 \nl
VLT/UVES     &  40\,000 & 3025--3918 & 1999  Oct  8,16   & LP815-43           & 10.91 &  10800 & 100 \nl
VLT/UVES     &  40\,000 & 3025--3918 &1999  Oct  9       &   G275-4           & 12.12 &  9000 & 80 \nl
VLT/UVES     &  55\,000 & 6652--10433& 2000  Feb 15      &   G64-12           & 11.47 &  9600 & 100 \nl
VLT/UVES     &  55\,000 & 3025--3918 & 2000 Feb   13,15  &   G64-12           & 11.47 & 20400 & 120 \nl  
\enddata
\tablenotetext{a}{The  signal-to-noise ratio (S/N) 
is provided at 7760 and 3130 \AA\ in
the red and blue spectra, respectively.} 
\end{deluxetable}


\clearpage

\begin{deluxetable}{lcccc}
\scriptsize
\tablecaption{Parameters from the literature and this work \label{tbl-3}}
\tablewidth{0pt}
\tablehead{
\colhead{Star}           & 
\colhead{$T_{\rm eff}$}  & 
\colhead{$\log g$}       & 
\colhead{[Fe/H]}         &
\colhead{Ref.}           \nl    
}
\startdata          
G4-37                 & 6050 &               & $-$2.7   & 2  \nl
                      & 6124 &               & $-$2.78  & 3  \nl
                      & 5837 & 4.23          & $-$2.89  & 5  \nl
G64-12                & 6380 & 4.39          & $-$3.12  & 1  \nl
                      & 6220 &               & $-$3.24  & 2  \nl
                      & 6197 &               & $-$3.52  & 3  \nl
                      & 6318 & 4.2           & $-$3.37  & 5  \nl
G64-37                & 6240 &               & $-$3.15  & 2  \nl
                      & 6376 &               & $-$2.70  & 3  \nl
                      & 6310 & 4.2           & $-$3.22  & 5  \nl
G268-32               & 5841 &               & $-$3.5   & 3  \nl
                      & 6000 & 3.8           & $-$2.7   & 4  \nl
                      & 6090 & 3.86          & $-$2.81  & 5  \nl
G271-162              & 5860 & 3.71          & $-$2.05  & 1  \nl
                      & 6135 &               & $-$2.69  & 3  \nl
                      & 6050 & 4.0           & $-$2.46  & 5  \nl 
G275-4                & 6000 & 4.39          & $-$3.12  & 1  \nl
                      & 6070 &               & $-$3.24  & 2  \nl
                      & 6168 &               & $-$3.70  & 3  \nl
                      & 6212 & 4.13          & $-$3.32  & 5  \nl
LP 815-43             & 6380 & 4.39          & $-$2.64  & 1  \nl
                      & 6340 &               & $-$3.00  & 2  \nl
                      & 6423 &               & $-$3.20  & 3  \nl
                      & 6265 & 4.54          & $-$3.05  & 5  \nl
LP 831-70             & 6000 & 4.40          & $-$2.85  & 1  \nl
                      & 6050 &               & $-$3.32  & 2  \nl
                      & 6119 &               & $-$3.40  & 3  \nl
                      & 6205 & 4.3           & $-$2.97  & 5  \nl
BD $+$03\arcdeg 740   & 6146 & 3.98          & $-$2.5   & 1  \nl
                      & 6240 &               & $-$2.7   & 2  \nl
                      & 6401 &               & $-$2.77  & 3  \nl
                      & 6135 & 4.0           & $-$2.93  & 5  \nl              

\enddata
\tablerefs{
(1): \cite{Thevenin+Idiart1999},
(2): \cite{Ryan+Norris+Beers1999},
(3): \cite{Thorburn1994},
(4): \cite{Norris+Ryan+Beers1997},
(5): present work.
}
\end{deluxetable}


\clearpage

\begin{landscape}
\begin{deluxetable}{lccccccccccc}
\scriptsize
\tablecaption{Abundances derived from the oxygen triplet and iron lines 
\label{tbl-4}}
\tablewidth{0pt}
\tablehead{
\colhead{Ion}           & 
\colhead{$\lambda$ (\AA)} &
\colhead{$\log gf$}  & 
\colhead{$\chi$}       & 
\colhead{G4-37}         &
\colhead{G64-37}      &
\colhead{G268-32}      & 
\colhead{G271-162}      &
\colhead{LP 831-70}      &
\colhead{BD $+$23\arcdeg 3130} &  
\colhead{BD $+$03\arcdeg 740}  \nl    
}
\startdata
O\,{\sc i}   & 7771.95  & 0.358    & 9.14 & 7.41 & $<$6.97 & 7.35 & 7.27 & $<$7.08 & 7.50 & 7.07  \nl
O\,{\sc i}   & 7774.18  & 0.212    & 9.14 & 7.43 &         & 7.21 & 7.31 &         & 7.48 & 7.01  \nl
O\,{\sc i}   & 7775.40  & $-$0.01 & 9.14 & 7.57 &         & 7.23 & 7.28 &         & 7.45 &       \nl
Fe\,{\sc i}  & 4271.774 & $-$0.173 & 1.49 &      & 4.24    & 4.61 & 5.24 & 4.55    & 5.26 &    \nl
Fe\,{\sc i}  & 4325.775 & 0.006    & 1.61 &      & 4.21    & 4.73 &      & 4.45    & 5.23 &    \nl
Fe\,{\sc i}  & 4383.557 & 0.208    & 1.48 &      & 4.23    & 4.79 &      & 4.45    & 5.27 &    \nl
Fe\,{\sc i}  & 4415.135 & $-$0.621 & 1.61 &      & 4.38    & 4.69 & 5.18 & 4.56    & 4.87 &    \nl
Fe\,{\sc i}  & 5227.192 & $-$1.227 & 1.56 & 4.61 & 4.26    & 4.73 &      & 4.49    & 5.10 & 4.59 \nl
Fe\,{\sc i}  & 5232.952 & $-$0.057 & 2.94 & 4.63 & 4.31    & 4.59 & 4.98 &         & 4.83 & 4.44 \nl
Fe\,{\sc i}  & 5269.550 & $-$1.333 & 0.86 & 4.62 & 4.31    & 4.67 &      & 4.66    & 4.98 & 4.62 \nl
Fe\,{\sc i}  & 5397.141 & $-$1.982 & 0.91 & 4.58 & 4.31    & 4.76 & 5.07 & 4.62    & 5.08 & 4.59 \nl
Fe\,{\sc i}  & 5405.785 & $-$1.852 & 0.99 & 4.61 & 4.35    & 4.77 & 5.01 & 4.62    & 5.06 & 4.58 \nl
Fe\,{\sc i}  & 5429.706 & $-$1.881 & 0.96 & 4.65 & 4.39    & 4.66 &      &         & 5.08 & 4.61 \nl
Fe\,{\sc i}  & 5434.534 & $-$2.126 & 1.01 & 4.61 & 4.35    & 4.80 & 5.02 &         & 5.09 & 4.57 \nl
Fe\,{\sc i}  & 5569.630 & $-$1.881 & 3.42 & 4.77 &         &      & 4.90 &         &      & 4.68 \nl
\hline
O(mean)  & & & & 7.47$\pm$0.09 &  &  7.26$\pm$0.07 & 7.29$\pm$0.03 & & 7.48$\pm$0.03 & 7.04$\pm$0.04 \nl        
Fe(mean) & & & & 4.63$\pm$0.06 & 4.30$\pm$0.06 & 4.71$\pm$0.07 & 5.06$\pm$0.12 & 4.55$\pm$0.08 & 5.07$\pm$0.14 
& 4.59$\pm$0.07 \nl  
\enddata
\end{deluxetable}
\end{landscape}


\clearpage

\begin{deluxetable}{lcccccccc}
\scriptsize
\tablecaption{Abundances derived from the OH and \FeI\ lines.  
\label{tbl-4}}
\tablewidth{0pt}
\tablehead{
\colhead{Ion}           & 
\colhead{$\lambda$ (\AA)} &
\colhead{$\log gf_{\rm th}$} &
\colhead{$\log gf_{\rm ad}$}  & 
\colhead{$\chi$}       & 
\colhead{G275-4}         &
\colhead{G64-12}      &
\colhead{LP815-43}      \nl    
}
\startdata
OH           & 3085.199 & $-$1.971 & $-$2.060 & 0.843 & 6.70   & 7.15   & 6.90 \nl
OH           & 3123.948 & $-$2.003 & $-$2.086 & 0.204 & 6.90   & 7.05   & 6.80 \nl
OH           & 3127.687 & $-$1.588 & $-$1.590 & 0.612 & 6.85   & 6.95   & 6.70 \nl
OH           & 3128.286 & $-$2.074 & $-$2.070 & 0.209 & 6.90   & 7.10   & 6.85 \nl
OH           & 3139.169 & $-$1.563 & $-$1.762 & 0.760 &        & 6.95   &        \nl
OH           & 3140.730 & $-$1.994 & $-$2.090 & 0.300 &        & 6.95   & 6.90 \nl
OH           & 3167.169 & $-$1.544 & $-$1.694 & 1.108 &        & 7.05   & 6.85 \nl
OH           & 3186.084 & $-$1.859 & $-$2.097 & 0.685 & 6.90   & 7.10   & 6.80 \nl
Fe\,{\sc i}  & 3786.682 & $-$2.185 &          & 1.01  & 4.26   & 4.23   & 4.43   \nl
Fe\,{\sc i}  & 3787.891 & $-$0.838 &          & 1.01  & 4.08   & 4.09   & 4.49   \nl
Fe\,{\sc i}  & 3790.098 & $-$1.739 &          & 0.99  & 4.23   & 4.21   & 4.53   \nl
Fe\,{\sc i}  & 3820.436 & 0.157    &          & 0.86  & 4.15   & 4.07   & 4.46   \nl
Fe\,{\sc i}  & 3821.187 & 0.198    &          & 3.27  &        & 4.20   & 4.52   \nl
Fe\,{\sc i}  & 3825.891 & $-$0.024 &          & 0.91  & 4.29   & 4.10   & 4.46   \nl
\hline
OH(mean)     &          &         & &       & 6.85$\pm$0.09 & 7.05$\pm$0.08 & 6.82$\pm$0.09 \nl
Fe(mean)     &          &         & &       & 4.20$\pm$0.08 & 4.15$\pm$0.07 & 4.48$\pm$0.04 \nl
\enddata
\tablenotetext{a}{The line $\lambda$ 3140.73 \AA\ has been added from the list of Nissen et al. (1994) and was
also used by Boesgaard et al. (1999).} 
\tablenotetext{b}{The $\log gf_{\rm th}$-values for 
the OH and \FeI\ lines are from Gillis et al. (2000) and O'Brian et al. (1991), respectively. 
$\log gf_{\rm ad}$ come from Israelian et al. (1998).}
\end{deluxetable}


\clearpage

\begin{deluxetable}{ccccr||ccccr}
\scriptsize
\tablecaption{\FeI\ line list for BD +23\arcdeg 3130
\label{tbl-5l}}
\tablewidth{0pt}
\tablehead{
\colhead{Wavelength}   & 
\colhead{$\chi$}       &
\colhead{$\log gf$}    & 
\colhead{Ref.}         & 
\colhead{Mult. no.}     &
\colhead{Wavelength }  &
\colhead{$\chi$}       & 
\colhead{$\log gf$}    &
\colhead{Ref.}         & 
\colhead{Mult. no.}     \nl
}
\startdata
4011.710 &    2.450  &  --2.693  &   4	&  153   &    5415.200 &    4.386  &   0.504  &   4   & 1165  \\
4055.030 &    2.559  &  --0.815  &   4	&  218   &    5424.070 &    4.320  &   0.576  &   4   & 1146   \\
4076.630 &    3.211  &  --0.356  &   4	&  558   &    5476.571 &    4.103  &  --0.842  &   4   & 1062  \\
4084.498 &    3.332  &  --0.597  &   4	&  698   &    5487.746 &    4.320  &  --0.650  &   5   & 1143   \\
4085.300 &    3.241  &  --0.708  &   4	&  559   &    5525.552 &    4.191  &  --1.080  &   3   & 1062   \\
4100.740 &    0.859  &  --3.179  &   1	&   18   &    5554.890 &    4.549  &  --0.444  &   4   & 1183   \\
4112.960 &    4.178  &  --0.325  &   4	& 1103   &    5576.097 &    3.430  &  --1.000  &   4   &  686   \\
4126.180 &    3.332  &  --0.959  &   4	&  695   &    5586.763 &    3.370  &  --0.100  &   3   &  686   \\
4181.750 &    2.832  &  --0.187  &   4	&  354   &    5615.652 &    3.332  &  --0.188  &   7   &  686   \\
4199.090 &    3.047  &   0.250  &   4	&  522   &    5662.525 &    4.178  &  --0.879  &   4   & 1087   \\
4216.186 &    0.000  &  --3.356  &   1	&    3   &    5701.550 &    2.560  &  --2.220  &   1   &  209   \\
4219.360 &    3.573  &   0.120  &   4	&  800   &    5934.660 &    3.929  &  --1.155  &   4   &  982  \\
4224.170 &    3.368  &  --0.400  &   4	&  689   &    5956.700 &    0.860  &  --4.605  &   1   &   14	\\
4276.680 &    3.882  &  --1.207  &   4	&  976   &    6003.030 &    3.882  &  --1.121  &   4   &  959   \\
4282.400 &    2.176  &  --0.810  &   4	&   71   &    6065.480 &    2.610  &  --1.530  &   1   &  207   \\
4285.440 &    3.237  &  --1.190  &   4	&  597   &    6136.610 &    2.450  &  --1.400  &   1   &  169  \\
4326.750 &    2.949  &  --1.920  &   4	&  413   &    6136.900 &    2.200  &  --2.950  &   1   &   62 \\
4388.410 &    3.603  &  --0.580  &   4	&  830   &    6151.620 &    2.180  &  --3.300  &   1   &   62   \\
4439.880 &    2.279  &  --3.002  &   4	&  116   &    6173.340 &    2.220  &  --2.880  &   1   &   62  \\
4445.470 &    0.087  &  --5.440  &   1	&    2   &    6200.320 &    2.610  &  --2.440  &   1   &  207   \\
4484.227 &    3.603  &  --0.724  &   4	&  828   &    6219.280 &    2.200  &  --2.430  &   1   &   62  \\
4528.610 &    2.176  &  --0.822  &   4	&   68   &    6230.730 &    2.560  &  --1.280  &   1   &  207  \\
4581.510 &    3.241  &  --1.833  &   4	&  555   &    6240.660 &    2.220  &  --3.230  &   3   &   64   \\
4602.944 &    1.485  &  --2.155  &   4	&   39   &    6246.334 &    3.600  &  --0.870  &   2   &  816  \\
4647.430 &    2.949  &  --1.350  &   4	&  409   &    6252.550 &    2.400  &  --1.690  &   1   &  169  \\
4733.596 &    1.490  &  --2.987  &   1	&   38   &    6265.130 &    2.180  &  --2.550  &   1   &   62   \\
4736.780 &    3.211  &  --0.752  &   2	&  554   &    6280.630 &    0.860  &  --4.390  &   1   &   13  \\
4957.603 &    2.808  &   0.043  &   4	&  318   &    6297.800 &    2.220  &  --2.740  &   1   &   62  \\
4966.096 &    3.332  &  --0.879  &   2	&  687   &    6301.515 &    3.650  &  --0.720  &   3   &  816  \\
4973.100 &    3.960  &  --0.955  &   4	&  984   &    6302.507 &    3.690  &  --1.150  &   6   &  816  \\
4978.600 &    3.984  &  --0.940  &   4	&  966   &    6322.690 &    2.590  &  --2.430  &   1   &  207   \\
5001.860 &    3.882  &  --0.004  &   4	&  965   &    6336.840 &    3.690  &  --0.860  &   3   &  816  \\
5012.071 &    0.860  &  --2.642  &   1	&   16   &    6358.690 &    0.859  &  --4.468  &   1   &   13 \\
\tablebreak
5044.210 &    2.850  &  --2.017  &   2	&  318   &    6393.600 &    2.430  &  --1.430  &   3   &  168   \\
5049.825 &    2.280  &  --1.355  &   2	&  114   &    6400.010 &    3.600  &  --0.290  &   3   &  816  \\
5162.270 &    4.178  &   0.019  &   2	& 1089   &    6411.658 &    3.650  &  --0.717  &   2   &  816 \\
5166.286 &    0.000  &  --4.195  &   1	&    1   &    6421.355 &    2.280  &  --2.027  &   1   &  111  \\
5171.600 &    1.485  &  --1.790  &   4	&   36   &    6430.851 &    2.180  &  --2.010  &   1   &   62 \\
5198.710 &    2.220  &  --2.135  &   1	&   66   &    6481.880 &    2.280  &  --2.980  &   1   &  109  \\
5202.339 &    2.180  &  --1.839  &   1	&   66   &    6494.980 &    2.390  &  --1.270  &   1   &  168  \\
5217.389 &    3.210  &  --1.162  &   2	&  553   &    6593.880 &    2.430  &  --2.420  &   1   &  168  \\
5232.946 &    2.940  &  --0.057  &   2	&  383   &    6609.120 &    2.560  &  --2.690  &   1   &  206  \\
5242.495 &    3.630  &  --0.967  &   2	&  843   &    6677.993 &    2.692  &  --1.418  &   2   &  268  \\
5247.052 &    0.087  &  --4.950  &   1	&    1   &    6750.150 &    2.420  &  --2.620  &   1   &  111  \\
5250.212 &    0.120  &  --4.940  &   1	&    1   &    6945.200 &    2.420  &  --2.480  &   1   &  111   \\
5253.460 &    3.280  &  --1.570  &   3	&  553   &    6978.850 &    2.480  &  --2.500  &   1   &  111 \\
5322.040 &    2.280  &  --3.022  &   4	&  112   &    7511.045 &    4.178  &   0.099  &   2   & 1077  \\
5324.185 &    3.210  &  --0.100  &   3	&  553   &    7748.281 &    2.949  &  --1.754  &   4   &  402 \\
5393.174 &    3.240  &  --0.720  &   3	&  553   &    7945.878 &    4.386  &   0.094  &   4   & 1154  \\
\enddata
\tablerefs{(1):
\cite{Blackwell+Ibbetson+Petford+Shallis1979},
\nocite{Blackwell+Petford+Shallis1979} 1979b,
\nocite{Blackwell+Petford+Shallis+Simmons1980} 1980,
\nocite{Blackwell+Petford+Shallis+Simmons1982} 1982a,
\nocite{Blackwell+Petford+Simmons1982} 1982b,
\nocite{Blackwell+Shallis+Simmons1982} 1982c,
\nocite{Blackwell+Booth+Petford1984} 1984,
\nocite{Blackwell+Booth+Haddock+Petford+Leggett1986} 1986,
\nocite{Blackwell+LynasGray+Smith1995} 1995:
(2): \cite{Obrian1991};
(3): \cite{Bard+Kock+Kock1991},
      \cite{Holweger+Bard+Kock+Kock1991},
       \nocite{Holweger+Kock+Bard1995} 1995;
(4) \cite{Fuhr+Martin+Wiese1988}.
}
\end{deluxetable}


\clearpage

\begin{deluxetable}{lcccc}
\scriptsize
\tablecaption{Abundances derived from \FeI\ and \FeII\ lines in four different models for BD +23\arcdeg 3130 \label{tbl-6}}
\tablewidth{0pt}
\tablehead{
\colhead{Abundance/model}           & 
\colhead{4850/2.0}           & 
\colhead{4825/2.5}           & 
\colhead{4980/2.7}           & 
\colhead{5130/3.0}           \nl      
}
\startdata
$\log \epsilon$ (\FeI)                  & 4.910$\pm$0.07  & 4.860$\pm$0.059 & 5.002$\pm$0.070 & 5.256$\pm$0.108 \nl
$\log \epsilon$ (\FeI) ($\chi < 3.8$eV) & 4.930$\pm$0.057 & 4.860$\pm$0.050 & 5.024$\pm$0.053 & 5.290$\pm$0.084 \nl
$\log \epsilon$ (\FeI) ($\chi > 3.8$eV) & 4.820$\pm$0.063 & 4.787$\pm$0.058 & 4.900$\pm$0.051 & 5.091$\pm$0.047 \nl
$\log \epsilon$ (\FeII)                 & 4.717$\pm$0.121 & 4.886$\pm$0.122 & 4.972$\pm$0.122 & 5.089$\pm$0.116 \nl
\enddata
\end{deluxetable}


\clearpage

\begin{landscape}
\begin{deluxetable}{lccccccccccc}
\scriptsize
\tablecaption{Stellar parameters and oxygen abundances in the sample of metal-poor 
unevolved stars \label{tbl-6}}
\tablewidth{0pt}
\tablehead{
\colhead{Star}                 & 
\colhead{$V-K$}                &
\colhead{$T_{\rm eff}$}        & 
\colhead{$\log g$}             & 
\colhead{[Fe/H]}         & 
\colhead{[Fe/H]}        &
\colhead{[O/H]$_{\rm Trip}$}   &
\colhead{[O/H]$_{\rm OH}$}     &
\colhead{[O/Fe]$_{\rm Trip}$}  &
\colhead{[O/Fe]$_{\rm Trip}$}  &
\colhead{[O/Fe]$_{\rm OH}$}       &
\colhead{[O/Fe]$_{\rm OH}$}    \nl
 & & & & LTE & NLTE$^{*}$ & & & LTE & NLTE$^{*}$ & LTE & NLTE$^{*}$ \nl      
}
\startdata
G4-37               & 1.450 & 5837$\pm$80  & 4.23$\pm$0.3  & $-$2.89$\pm$0.1  & $-$2.59  & $-$1.51$\pm$0.16    &          & 1.38     &  1.08 &    &  \nl
G64-12              & 1.196 & 6318$\pm$150 & 4.20$\pm$0.3  & $-$3.37$\pm$0.16 & $-$3.05 & $-$2.14    & $-$1.88$\pm$0.31 & 1.23   & 0.91   & 1.49 & 1.17 \nl
G64-37              & 1.191 & 6310$\pm$110 & 4.20$\pm$0.3  & $-$3.22$\pm$0.12 & $-$2.90 & $<-$2.01            &         & $<$1.21   & $<$0.89   &  &   \nl
G268-32             & 1.306 & 6090$\pm$100 & 3.86$\pm$0.4  & $-$2.81$\pm$0.12 & $-$2.51 & $-$1.72$\pm$0.18    &          & 1.09     &  0.79   &    &     \nl
G271-162            & 1.320 & 6050$\pm$70  & 4.03$\pm$0.3  & $-$2.46$\pm$0.14 & $-$2.18 & $-$1.69$\pm$0.16    &         & 0.77      & 0.5    &  &    \nl
G275-4              & 1.254 & 6212$\pm$150 & 4.13$\pm$0.3  & $-$3.32$\pm$0.17 & $-$2.99 & $<-$1.96   & $-$2.08$\pm$0.32 & $<$1.36 & $<$1.03  & 1.24 & 0.91 \nl
LP 815-43           & 1.218 & 6265$\pm$125 & 4.54$\pm$0.3  & $-$3.05$\pm$0.13 & $-$2.74 &            & $-$2.11$\pm$0.27 &      &  &  0.94    & 0.63 \nl
LP 831-70           & 1.251 & 6205$\pm$120 & 4.30$\pm$0.3  & $-$2.97$\pm$0.14 & $-$2.66 & $<-$1.90   &                  & $<$1.07  & $<$0.76 &  &      \nl
BD $+$23\arcdeg 3130 & 1.97 & 5130$\pm$150 & 3.05$\pm$0.25 &      -           & $-$2.43  & $-$1.50$\pm$0.14    & $-$1.45$\pm$0.33  &          & 0.93  &      &  0.98  \nl
BD $+$03\arcdeg 740  & 1.284 & 6135$\pm$100 & 4.00$\pm$0.1 & $-$2.93$\pm$0.12 & $-$2.63 & $-$1.94$\pm$0.13    &         & 0.99    & 0.69  &     &      \nl
\tablenotetext{}{NLTE$^{*}$ refers to the [Fe/H] ratio estimated after Th\'evenin \& Idiart (1999).
See section 3 for details. The NLTE Fe abundance of $\rm BD +23\degree 3130$ is estmiated in this paper.} 
\enddata
\end{deluxetable}
\end{landscape}

\clearpage

\begin{figure}
\psfig{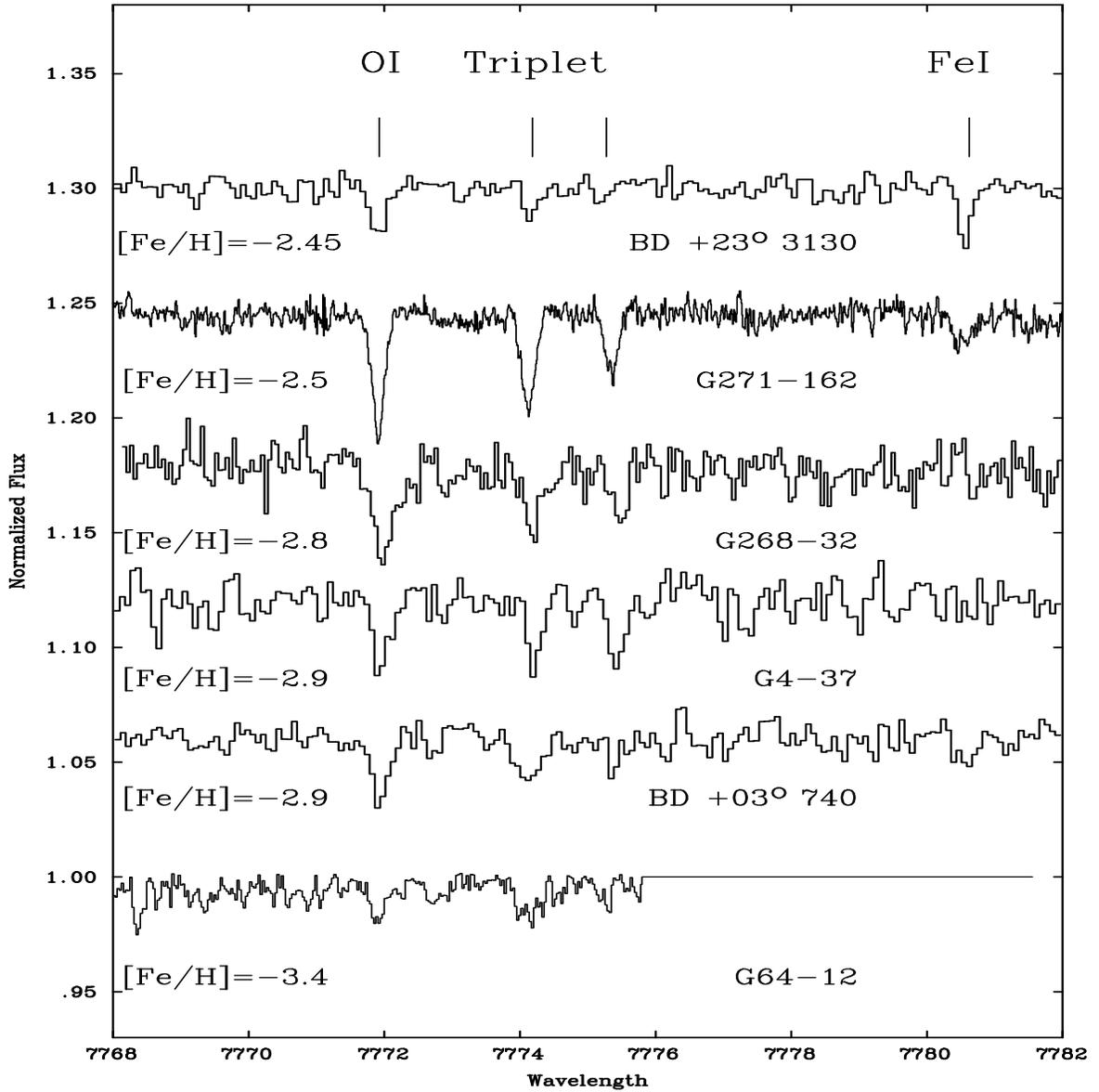}
\caption[]{Oxygen triplet observed in the metal-poor subdwarfs. The
metallicities quoted have been estimated in LTE.}
\end{figure}


\begin{figure}
\psfig{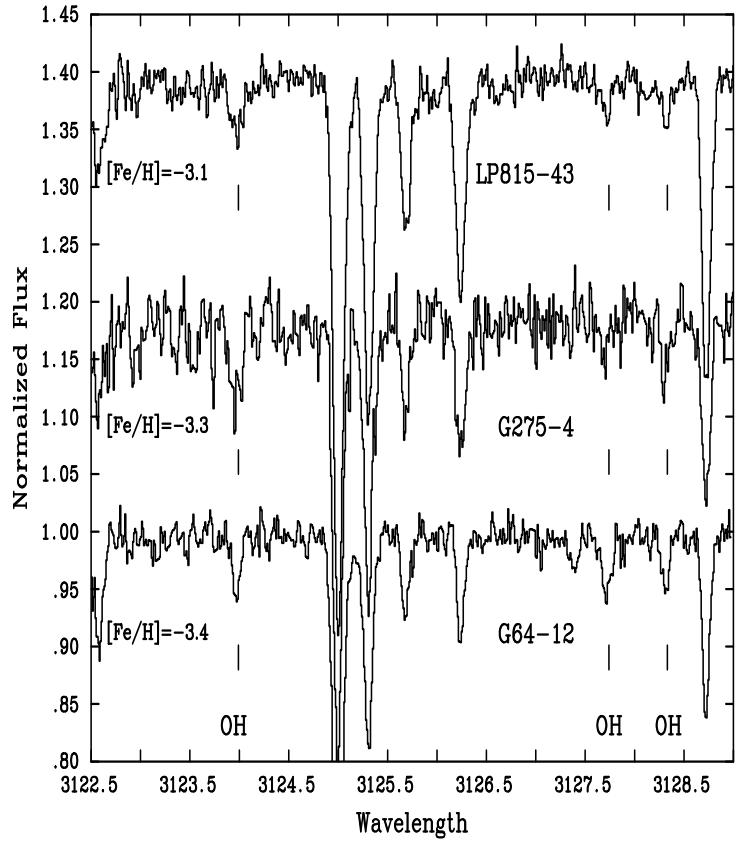}
\caption[]{OH lines in the UVES spectra of three ultra-metal-poor
subdwarfs.  The metallicities quoted have been estimated in LTE.}
\end{figure}


\begin{figure}
\psfig{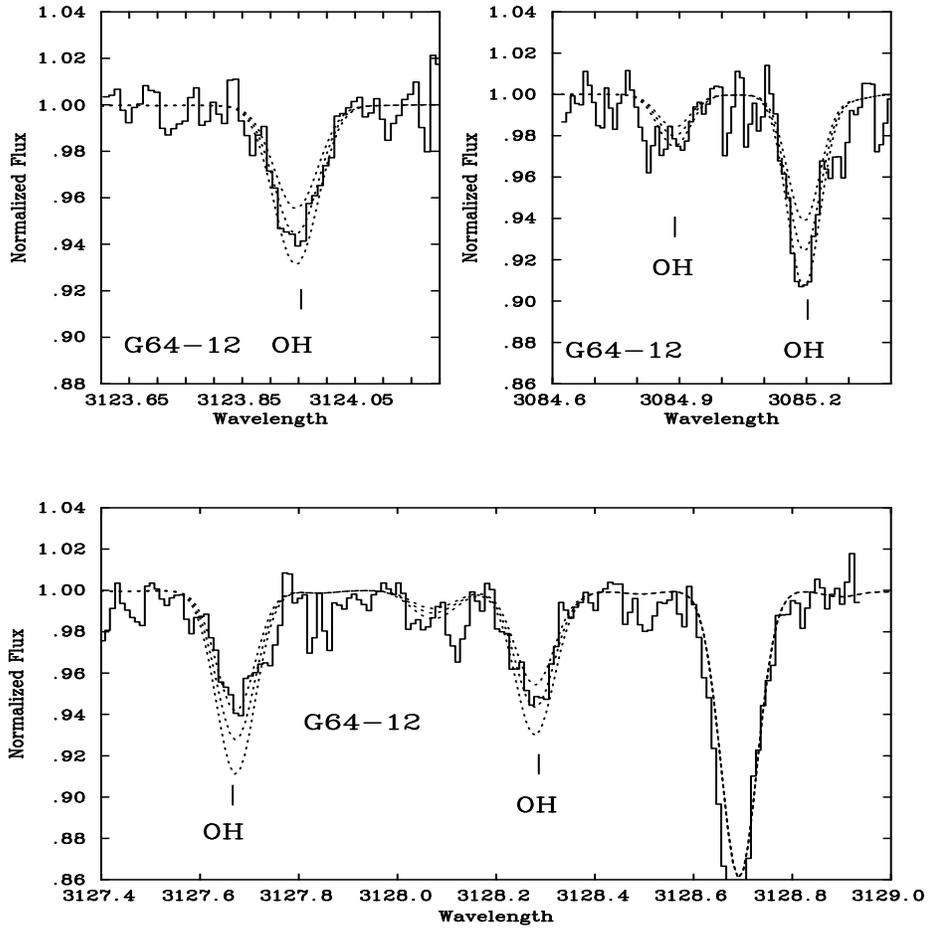}
\caption[]{Synthetic spectra fits to the OH lines in the UVES
spectrum of G64-12. The synthetic spectra are computed for 
$\log \epsilon$(O)=6.95, 7.05 and 7.15.}
\end{figure}


\begin{figure}
\psfig{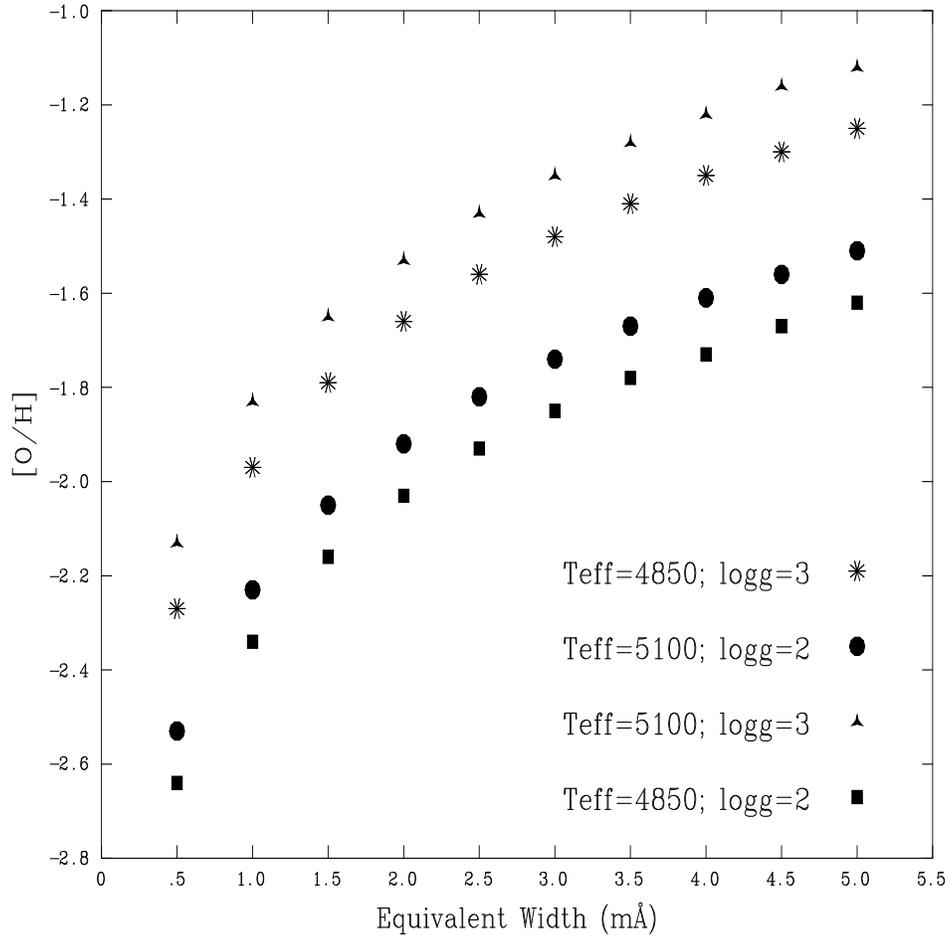}
\caption[]{The oxygen abundance derived from the forbidden line [O\,{\sc i}] 6300 \AA\
for different values of the equivalent widths and four atmospheric models.}
\end{figure}


\begin{figure}
\psfig{figure=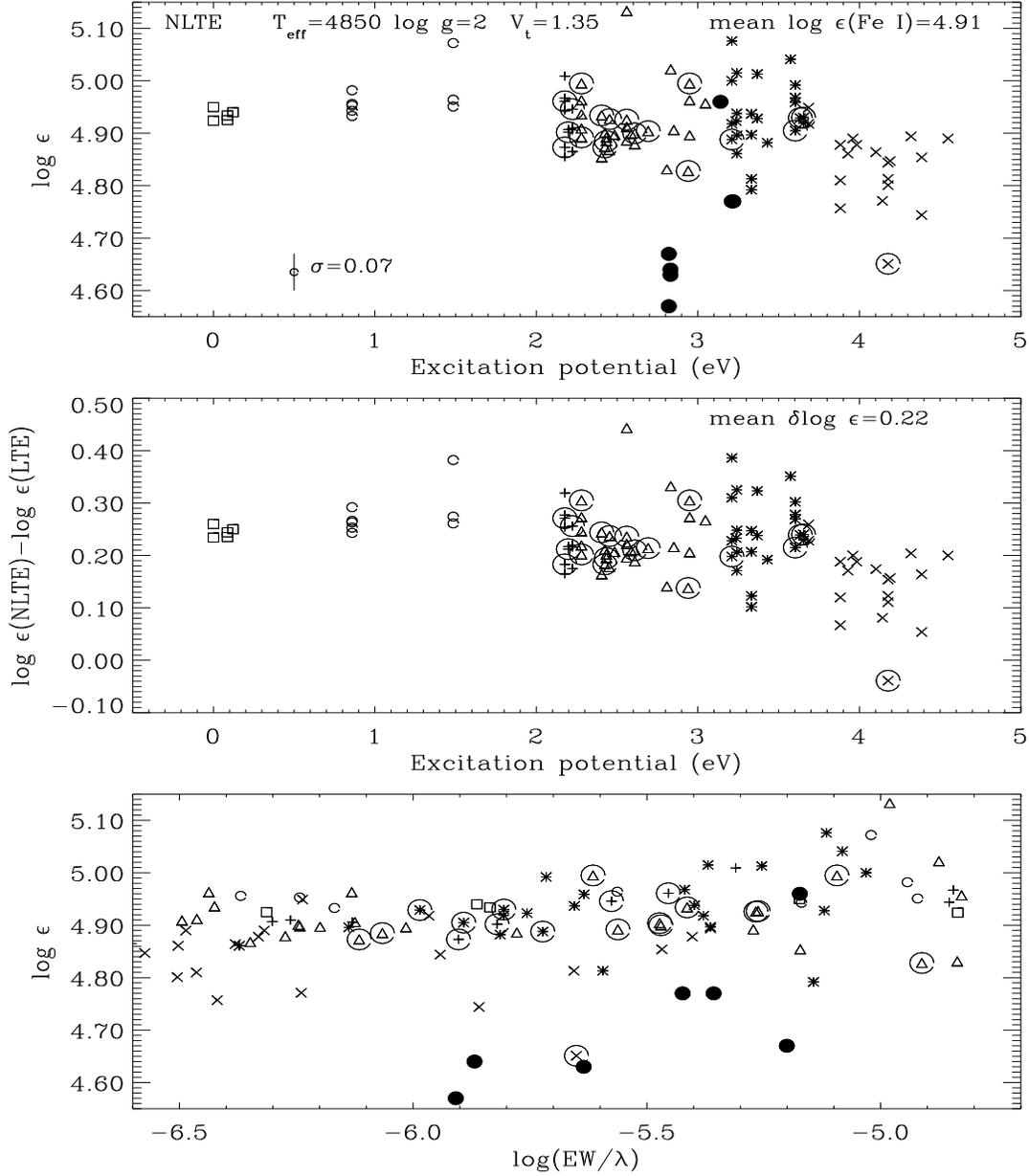,width=15.0cm,height=16.5cm,angle=360}
\caption[]{ Results of NLTE abundance determinations for \FeI\ lines
               and model atmosphere with parameters $T_{\rm eff}=4850$ K, 
               $\log g=2.0$, 
               $V_t=1.35$, 
               $\log\epsilon = 4.69$. The lines with $\chi < $ 0.8 eV are
marked as squares, the lines with $0.8 < \chi < $ 1.8 eV are marked as small unfilled
circles, the lines with $1.8 < \chi < $ 2.2 eV are marked with pluses, 
$2.2 < \chi < $ 3.0 eV triangles, $3 < \chi < $ 3.8 eV are stars, 
$3.8 < \chi < $ 4.8 eV are crosses. The lines of \FeII\ are marked
as filled circles. 
   {\it Top:}   NLTE abundances as a function of the lower excitation
               potential $\chi$.
{\it Middle:}  NLTE abundance corrections \vs $\chi$.
{\it Bottom:}  NLTE abundances as a function
               reduced equivalent widths $\log (EW{/}\lambda)$.
               The results for \FeI\ lines of different 
               $\chi$-classes are denoted by different symbols.
               \FeI\ lines from the line list by
               \cite{FK1999} are marked by large unfilled
               circles.}
\label{fig:abu_fk}
\end{figure}


\begin{figure}
\psfig{figure=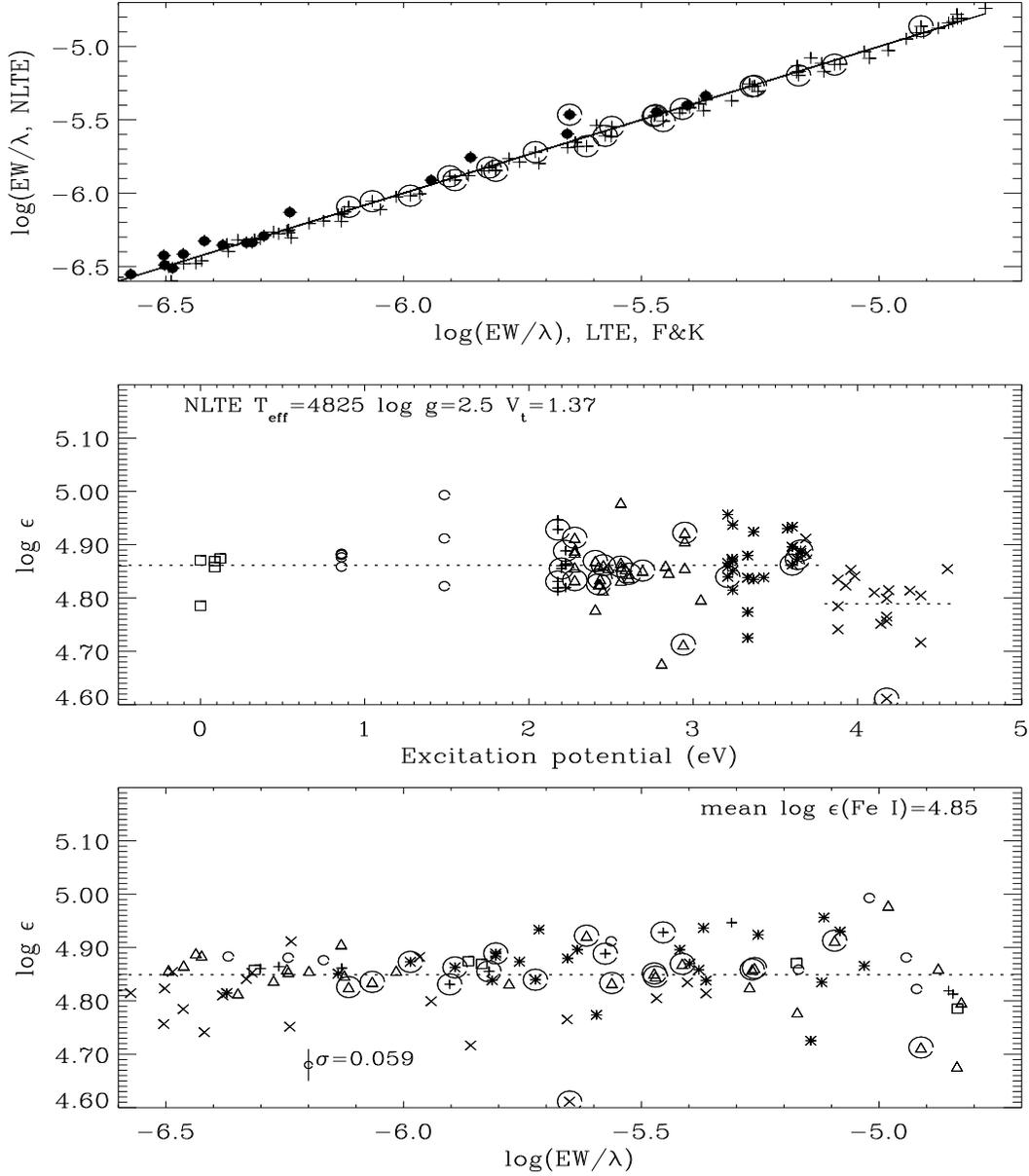,width=15.0cm,height=16.5cm,angle=360}
\caption[]{Computed equivalent widths $\log EW{/}\lambda$ 
              and NLTE abundances $\log \epsilon$ 
              for \FeI\ lines and 
              for the parameters of the 
              star $\rm BD +23\degree 3130$:
              $T_{\rm eff}=4825$ K, $\log g=2.5$, $V_t=1.37$.
   {\it Top:} Computed NLTE equivalent widths for \FeI\ 
              lines versus ``quasi-observed'' ones. 
              The latter are calculated for the parameters 
              found by Fulbright \& Kraft (1999)
              using an LTE approach
              ($T_{\rm eff}=4850$ K, 
               $\log g=2.0$, 
               $V_t=1.35$, 
               $\log\epsilon = 4.69$).
               The results for the high excitation  lines
               are denoted by filled circles.
 {\it Middle:} NLTE abundances as a function of 
               the lower excitation potential $\chi$.
 {\it Bottom:} The same values against $\log (EW{/}\lambda$.)}
\label{fig:best_wa}
\end{figure}

\begin{figure}
\psfig{figure=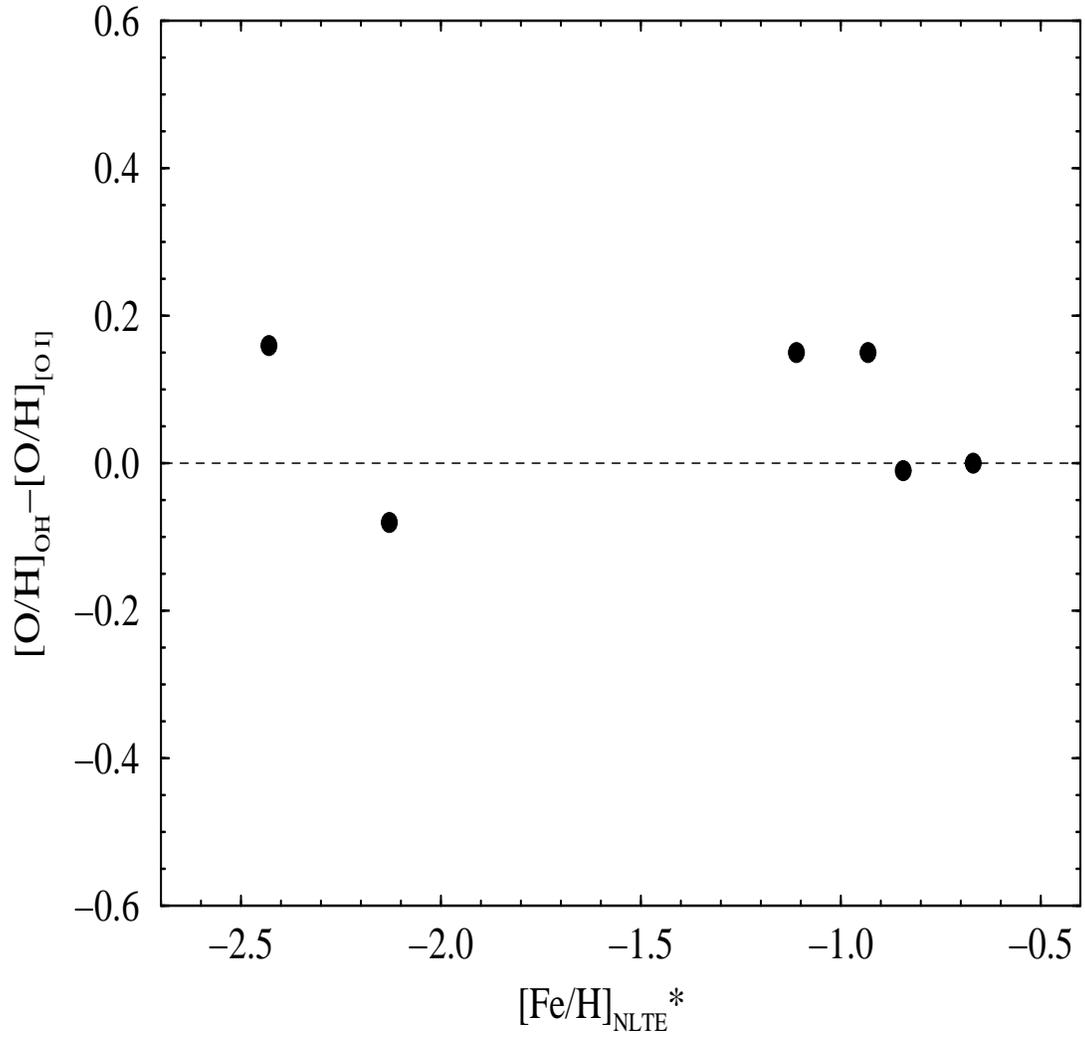,width=17.5cm,height=15.cm,angle=270}
\caption[]{
Differences between oxygen abundances derived from OH and forbidden
lines for two stars discussed by FK and four stars from Israelian et al. (1998).
The iron abundances have been corrected for NLTE effects as described in section 3. 
}
\end{figure}


\begin{figure}
\psfig{figure=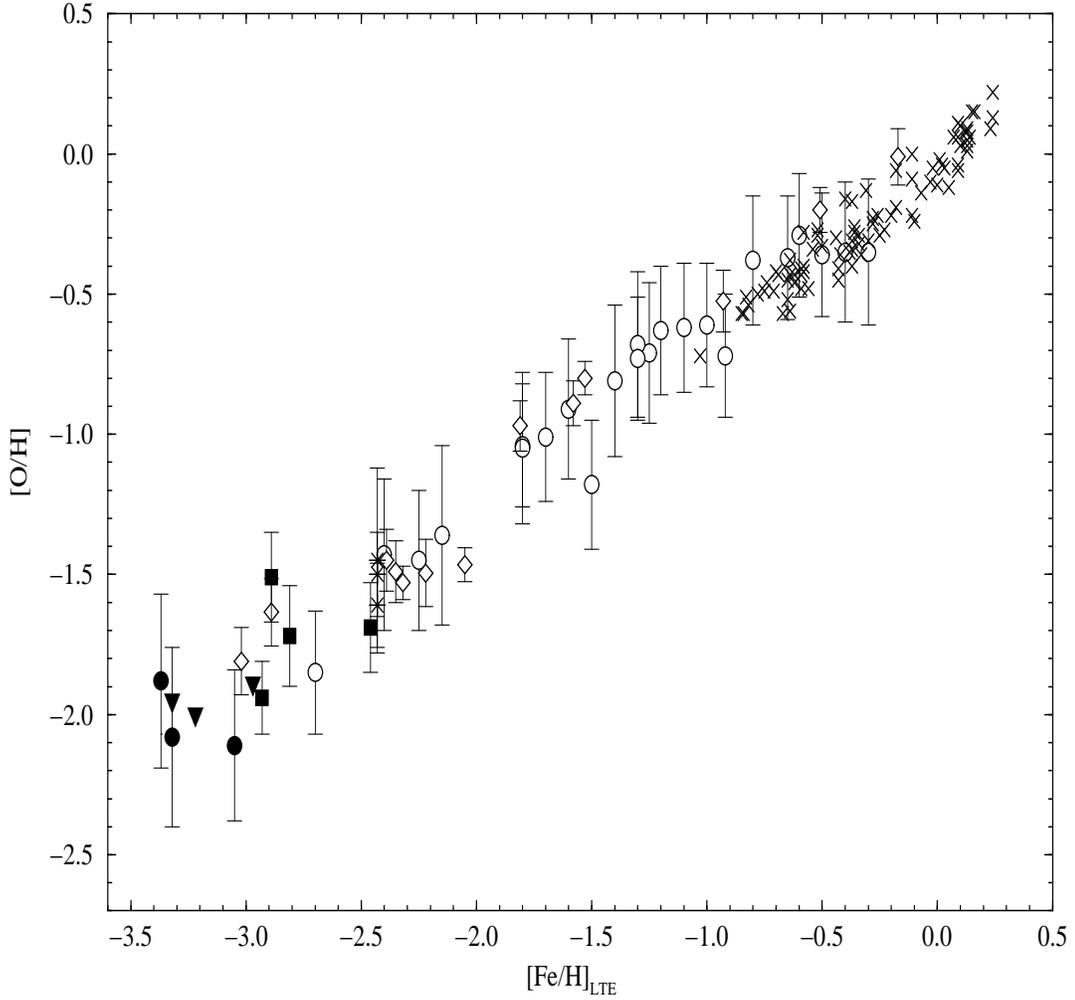,width=17.5cm,height=15.cm,angle=270}
\caption[]{
New oxygen abundances derived in this work
together with those measured by Israelian et al., Boesgaard et al. and
Edvardsson et al. (1993). The results for OH lines are denoted by 
filled circles, triplet by filled squares, and upper 
limits from triplet by filled triangles. The asterisks indicate the oxygen 
abundances derived from OH, triplet and [O\,{\sc i}] for $\rm BD +23\degree 3130$.
Data from the papers of Israelian et al. (1998), Boesgaard et al. (1999a) and 
Edvardsson et al. (1993) are marked by unfilled circles, unfilled 
diamonds and crosses, respectively.
}
\end{figure}


\begin{figure}
\psfig{figure=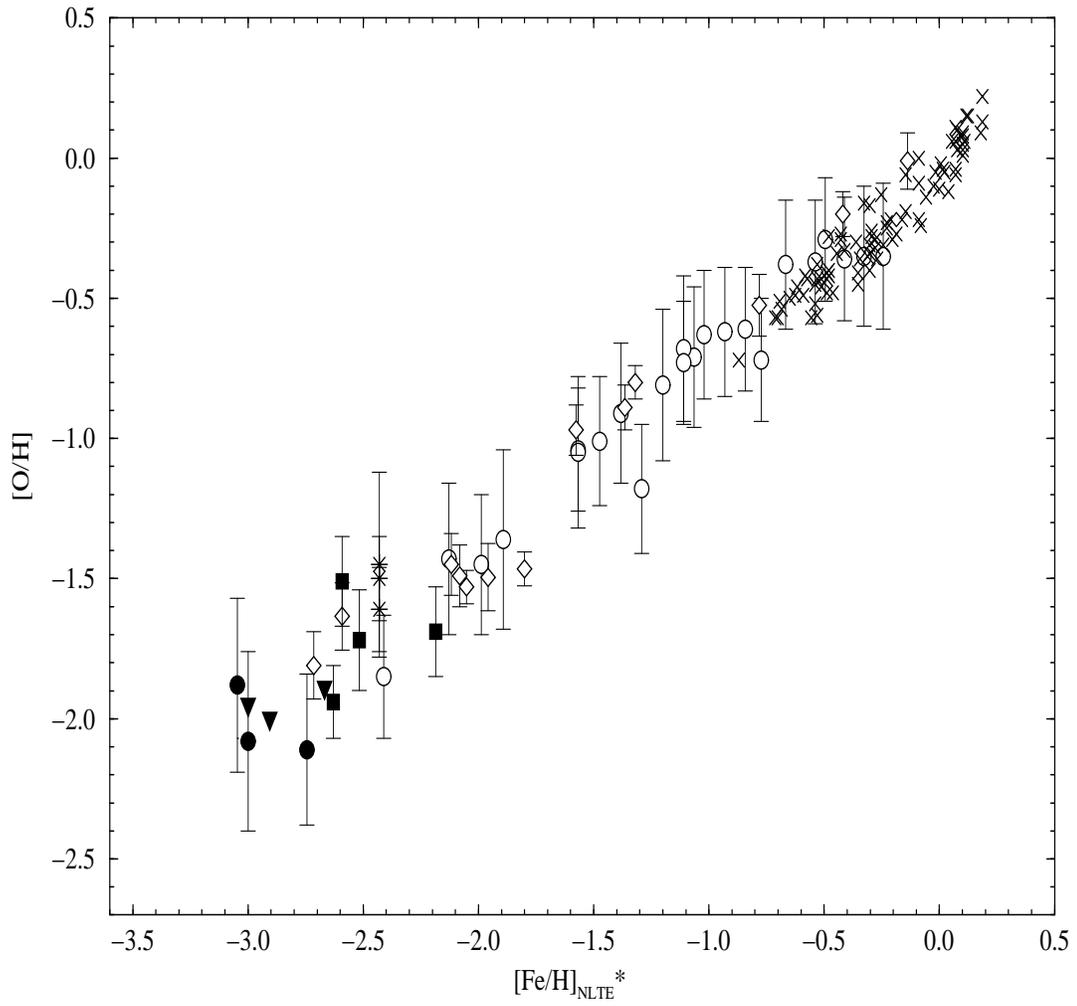,width=17.5cm,height=15.cm,angle=270}
\caption[]{
The same as in Fig 8. but with NLTE corrections estimated for [Fe/H]. 
}
\end{figure}


\begin{figure}
\psfig{figure=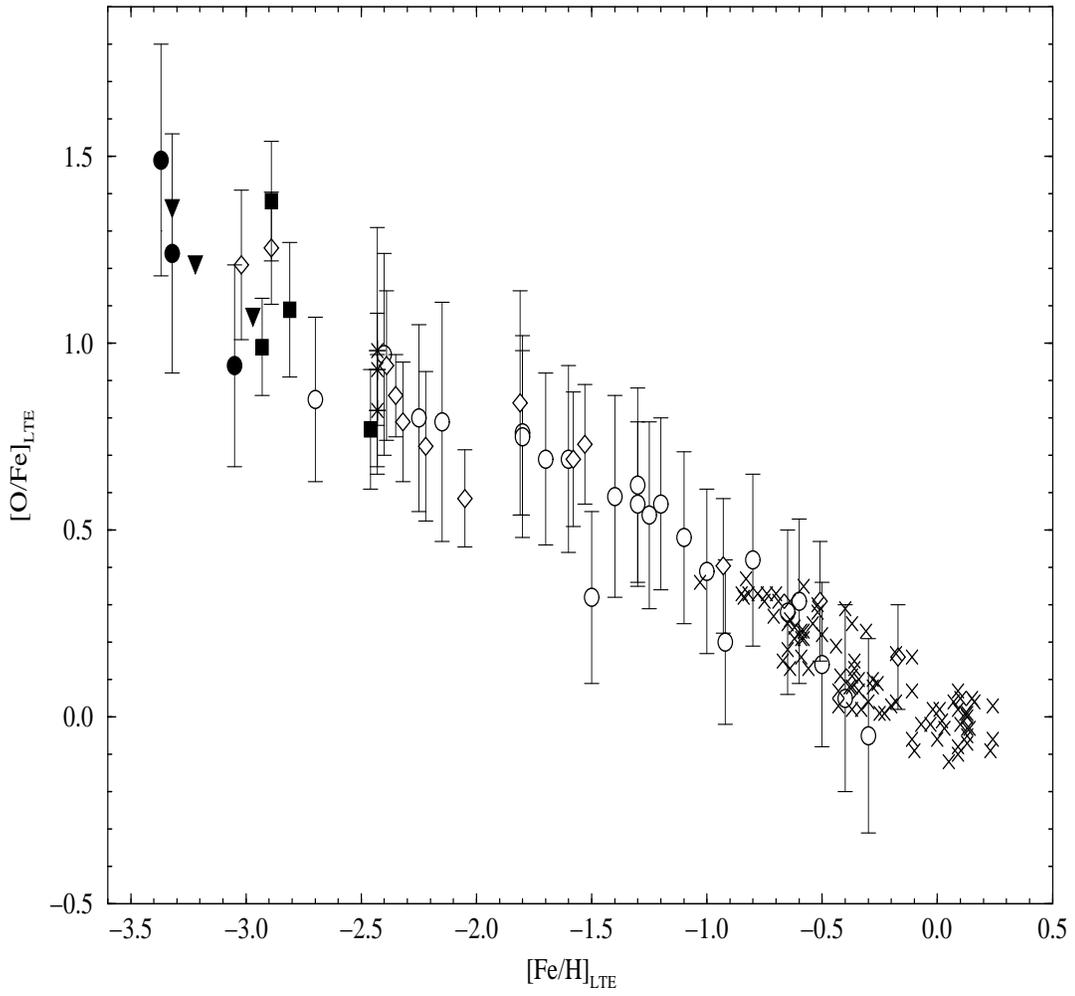,width=17.5cm,height=15.cm,angle=270}
\caption[]{Oxygen overabundance with respect to iron, derived from the UV OH lines 
and the O\,{\sc i} triplet in metal-poor unevolved stars. 
Symbols as in Fig 8.}
\end{figure}


\begin{figure}
\psfig{figure=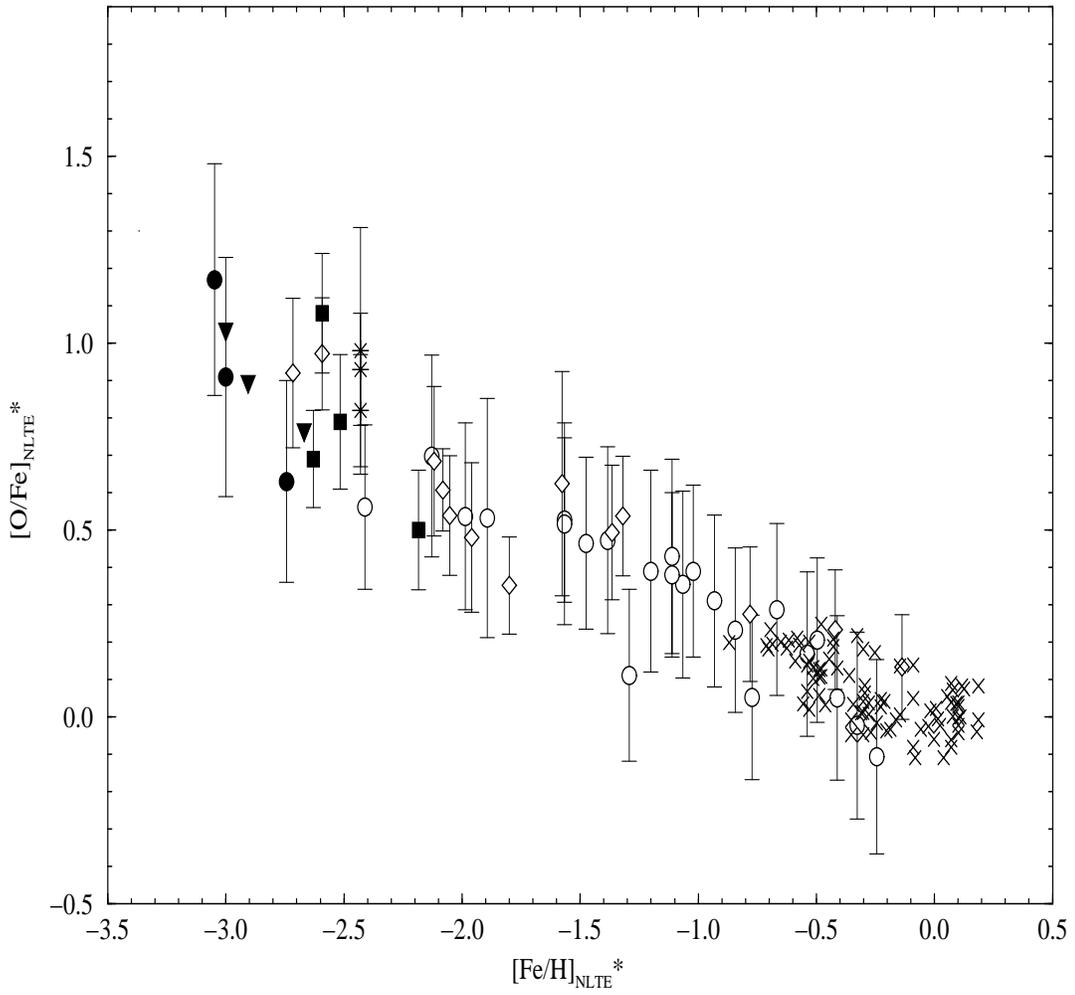,width=17.5cm,height=15.cm,angle=270}
\caption[]{The same as in Fig 10 but with NLTE estimates of [Fe/H].}
\end{figure}


\end{document}